\documentclass[sn-basic]{sn-jnl}% Basic Springer Nature Reference Style/Chemistry Reference Style
\usepackage{graphicx}
\usepackage{array}
\usepackage{multirow}%
\usepackage{amsmath,amssymb,amsfonts}
\usepackage{amsthm}
\usepackage{mathtools}
\usepackage{mathrsfs}
\usepackage[title]{appendix}%
\usepackage{xcolor}%
\usepackage{textcomp}%
\usepackage{manyfoot}%
\usepackage{booktabs}%
\usepackage[linesnumbered,ruled]{algorithm2e}
\usepackage{algorithmicx}
\usepackage{algpseudocode}
\usepackage{listings}
\usepackage{comment}
\theoremstyle{thmstyleone}%
\theoremstyle{thmstyletwo}%

\theoremstyle{thmstylethree}%

\begin{document}

\title{An Experimental Study of Existing Tools for Outlier Detection and Cleaning in Trajectories}

\author*[1]{\fnm{Mariana}  \sur{M Garcez Duarte}}\email{mariana.machado.garcez.duarte@ulb.be}

\author*[1,2]{\fnm{Mahmoud} \sur{Sakr}}\email{mahmoud.sakr@ulb.be}

\affil[1]{ \orgname{Université libre de Bruxelles}, \orgaddress{\city{Brussels}, \country{Belgium}}}

\affil[2]{\orgname{Ain Shams University}, \orgaddress{\city{Cairo}, \country{Egypt}}}


\abstract{Outlier detection and cleaning are essential steps in data preprocessing to ensure the integrity and validity of data analyses. This paper focuses on outlier points within individual trajectories, i.e., points that deviate significantly inside a single trajectory. We experiment with ten open-source libraries to comprehensively evaluate available tools, comparing their efficiency and accuracy in identifying and cleaning outliers. This experiment considers the libraries as they are offered to end users, with real-world applicability. We compare existing outlier detection libraries, introduce a method for establishing ground-truth, and aim to guide users in choosing the most appropriate tool for their specific outlier detection needs. Furthermore, we survey the state-of-the-art algorithms for outlier detection and classify them into five types: Statistic-based methods, Sliding window algorithms, Clustering-based methods, Graph-based methods, and Heuristic-based methods. Our research provides insights into these libraries' performance and contributes to developing data preprocessing and outlier detection methodologies.
}


\maketitle
\section{Introduction}

 Trajectory data refers to the path that a moving object follows through space as a function of time and finds extensive applications in various fields such as transportation, logistics, and environmental studies. However, outliers or noise often affect this data type due to sensor errors and connectivity issues.  Outliers are data points that deviate significantly from the expected pattern. In the context of trajectory data, these could represent an abrupt and uncharacteristic change in direction or speed. Such outliers can yield misleading analysis results and inaccurate decision-making, impacting businesses, governments, and individuals. For instance, in transportation, outliers in vehicle trajectory data could lead to incorrect traffic flow analysis, resulting in inefficient traffic management. This makes outlier or noise detection a crucial step in trajectory cleaning \cite{MobilityDB4}. As such, outlier detection is an essential function in mobility data management systems \cite{MobilityDB,MobilityDB2}.

There are two outlier detection categories, one focusing on a collection of trajectories, where a trajectory can be an outlier, and another on points inside one trajectory. In this paper, we focus on the latter.

{\color{blue} 
This paper is an extension of \cite{DuarteS23}. While the previous paper focused on the state of technology, this extension additionally surveys the state of the art in outlier detection algorithms, which is not limited to the algorithms implemented in the surveyed libraries. As such, the extension covers both the state of the art and the state of the tools. Additionally, an experimental study of libraries Scikit-learn \cite{scikit-learn} and MvOutlier \cite{mvoutlier} focuses on outlier detection. This study considers the user's perspective. That is, the context is to compare the offering of the existing libraries for end users rather than comparing their algorithms and implementation aspects. Finally, we have implemented a heuristic method for outlier detection in the MEOS library \cite{MobilityDB} and added it to the experiments.
}

\textit{Outline.} The rest of the paper is structured as follows. In Section \ref{sec:outlier} we present essential concepts for outlier detection. Section \ref{sec:stateoftheart} surveys the state of art. An experimental study using real data is the subject of Section \ref{sec:experiments}. In Section \ref{sec:conclusion}, we discuss our findings, and we conclude.

\section{Outlier Detection}
\label{sec:outlier}

There are a variety of outlier detection techniques. In \cite{newSurvey}, the authors divide the methods into five categories:  Statistical-based methods, Sliding window algorithms, Clustering-based methods, and Graph-based methods. We present here a brief survey, by no means exhaustive, of outlier detection techniques.

\paragraph{Statistic-based methods} 
The essential concept of these techniques for identifying outliers relies on their alignment with the statistical distribution model. These methods include the Kalman filter (KF). KF is a well-established method used to smooth point series. This algorithm estimates missing points based on previously observed values that might have measurement errors. In \cite{kf60}, the authors mention the advantage of the KF and its derivatives is its recursive aspect, which can be used in real time. It is also widely used due to its simplicity and capability to provide accurate estimations and prediction results.

 Particle filters (PF) use a set of randomly generated particles to represent the possible states of the system and update the particles based on observed data. In contrast to KF, they are not restricted to the Gaussian distribution of errors, which makes them applicable to a broader range of noisy data. PF can, however, be computationally intensive and thus not commonly implemented in trajectory libraries. Additionally, like KF, PF is sensitive to the first measurement in the trajectory, and their accuracy can be reduced if the first point is an outlier \cite{SurveyTrajectoryDM, particleFilter, GparticleFilter}. 

The Hampel Filter (HF) detects and replaces outliers in trajectories with estimates via the Hampel identifier. The HF expresses a conventional heuristic that almost all values lie within three standard deviations of the mean \cite{hampelTheory}. For each trajectory, the method calculates the median of a sliding window and adjacent points on each side of the trajectory. The HF estimates the standard deviation of each point about its window median using the median absolute deviation. If a measurement differs from the median by more than the threshold, the filter replaces the sample with the median.

\paragraph{Sliding window algorithms} 
The foundation of sliding window algorithms is calculating the distance between different data points in their neighborhood. An outlier is a data point that is significantly distant from its neighbors. When detecting an outlier inside a trajectory, the method compares each point in the trajectory to its neighbors and selects the points significantly further away than expected. Methods based on the mean or Median Filters replace points compared to the measurements done at preceding points in time. These algorithms are simple and practical for detecting single outliers.
Nonetheless, these techniques depend on the number of predecessors compared to the mean. Multiple successive outlier points can affect the accuracy of the outcome trajectory. These algorithms are only sometimes effective at detecting multiple successive outlier points, which can affect the accuracy of the output trajectory. More advanced algorithms or methods may be needed to detect and correct outliers in these cases.

In \cite{Distance-BasedOutliers}, the authors review a series of sliding window algorithms. They develop an optimized cell-based algorithm that outperforms the existing methods. The cell-based methodology is built on partitioning the multidimensional data space into a structured grid of cells. The dimensions of these cells are contingent on the distance threshold established for detecting outliers. Subsequently, each cell within the grid is searched to identify potential outliers. In experiments, the method exhibits scalability with large datasets. However, the computational cost escalates significantly as the dimension increases. The optimization of the cell-based methodology entails a strategic approach to minimize the number of cells that necessitate examination. This is accomplished by maintaining a register of active cells containing at least one object. Only these active cells are subject to examination in the ensuing passes over the dataset.

{\color{blue}  In \cite{newslidingWindow}, the authors tackle identifying outliers within extensive trajectory streams. The paper introduces a taxonomy of neighbor-based trajectory outlier definitions. The authors evaluate their work on two real-world datasets, the Beijing Taxi trajectory data and the Ground Moving Target Indicator (GMTI) data stream. The authors also utilize the synthetic dataset Moving Objects Database (MOD) generated by a network-based moving object generator \cite{datasetwi} using real road networks. The Minimal Examination (MEX) framework is central to their approach. It embodies three optimization principles such as point neighbor–based outliers (PN-Outliers), trajectory neighbor–based outliers (TN-Outliers), and synchronized neighbor–based outliers (SN-Outliers) that exploit spatiotemporal and predictability properties of neighbor evidence to reduce detection costs significantly. The framework has algorithms that detect the outliers based on these classes of new outlier semantics. The testing shows that MEX is scalable and can handle up to one million moving objects per second on a desktop.

The paper \cite{window2} introduces a method for outlier detection in trajectory data streams, proposing two algorithms: trajectory outlier detection on trajectory data Streams (TODS) and an approximate trajectory outlier detection (ATODS) algorithm aimed at reducing detection costs by a space approximate approach. The effectiveness of these methods is validated through experiments with both real and synthetic data, demonstrating their capability to maintain accuracy while reducing computational load.

The paper \cite{Shi2021RUTODRU} presents a framework, RUTOD, developed for the real-time detection of outliers in urban traffic data streams. Utilizing Apache Flink, RUTOD integrates Individual Outlier Detection (IOD) and Group Outlier Detection (GOD) to process streaming data efficiently. It employs a street-based grouping method for data preprocessing, feature-based outlier detection for IOD, and a density-based Local Outlier Factor (LOF) algorithm for GOD. The processing of incoming trajectory streams is organized within each time window into micro-batches. Experiments indicate a superior performance in throughput and latency compared to existing methods while maintaining comparable accuracy.
}

\paragraph{Clustering-based methods}
 Clustering-based techniques leverage standard clustering algorithms to distinguish outliers in the data. These techniques leverage the data's inherent structure and density characteristics to differentiate outliers from regular data points. In these methods, data points that do not belong to or lie close to large or dense clusters are considered outliers.

One such algorithm employs this approach is DBSCAN (Density-Based Spatial Clustering of Applications with Noise) \cite{DBSCAN}. DBSCAN is a density-based clustering algorithm that identifies clusters as high-density regions separated by regions of lower density. It is particularly effective in detecting outliers as it does not require the specification of the number of clusters, and it can discover clusters of arbitrary shape, unlike many other clustering algorithms. Outliers in DBSCAN are identified as points that do not belong to any cluster.

Another algorithm that utilizes clustering-based techniques for outlier detection is CLARANS (Clustering Large Applications based upon RANdomized Search) \cite{clarans}. CLARANS is an algorithm for medoid (representative objects) based clustering, a partitioning-based clustering variant. The algorithm identifies outliers as points far from the medoids of any clusters. CLARANS has the advantage of being more robust to noise and outliers than other partitioning-based clustering algorithms.

BIRCH (Balanced Iterative Reducing and Clustering using Hierarchies) \cite{BIRCH} is another algorithm that relies on clustering-based techniques. BIRCH is a hierarchical clustering algorithm that can incrementally and dynamically cluster incoming multi-dimensional data points to produce the best quality clustering with the memory available and time constraints. In the context of outlier detection, BIRCH identifies outliers while building the CF (Clustering Feature) tree, a data structure used to summarize the information of data points in the dataset. Data points that do not fit well into the structure of the CF tree are considered outliers.

K-Means \cite{kmeans} is a partitioning method that divides n observations into k clusters, each belonging to the nearest mean cluster. The algorithm operates iteratively and randomly selects `k' data points from the dataset. These points act as the initial partitions of the clusters. A Euclidean distance to each centroid is computed for the remaining data points, and the point is assigned to its nearest centroid. Once all points have been assigned to clusters, the partitions of the clusters are recalculated. This process continues until the partitions no longer move significantly or a set number of iterations is reached. K-mean is an efficient and straightforward method, particularly suitable for large datasets. However, the algorithm has a few limitations with complex geometrically distributed data. It is also sensitive to the initial choice of partitions and can fall into local minima. Furthermore, the number of clusters `k' needs to be specified beforehand, which can be a drawback if the data does not suggest a precise number of clusters.

{\color{blue} 
Clustering-based methods could work with long trajectories with possibly sparse or irregularly sampled locations. For example, in \cite{Eldawy2020}, the algorithm partitions a long trajectory into line segments and summarizes those segments without affecting the spatial properties of the original trajectory. In addition, in \cite{math11030620} the authors propose the Crowdsourcing Trajectory Outlier Detection (CTOD), which employs an adaptive spatial clustering algorithm to remove location offset points in trajectory sequences. Additionally, the paper introduces a 6-dimensional movement feature vector for each trajectory point. It utilizes a Temporal Convolutional Network Autoencoder (TCN-AE) with a Squeeze-and-Excitation (SE) channel attention mechanism to enhance detection. }

\paragraph{Graph-based methods}
   Graph-based techniques are a set of methodologies that leverage the structure of graphs to capture the relationships between interconnected data points. These techniques are effective in identifying outliers. Points are constituted as nodes, and their relationships are represented as edges. The weight of an edge typically represents the similarity or distance between two data points. The graph thus formed encapsulates the data structure, and the properties of this graph are used to identify outliers. Outliers are often identified as nodes that have few connections or are connected to other nodes with weak relationships. In other words, outliers are typically data points that do not fit well into the graph structure.

One common graph-based technique for outlier detection is the Spectral Clustering algorithm \cite{ng2002spectral}. This algorithm uses the eigenvalues of the Laplacian matrix of the graph, a matrix representation of the graph, to identify clusters of nodes. Outliers are often found as nodes that do not belong to these clusters. 

Another graph-based technique is the Local Outlier Factor (LOF) algorithm \cite{LOF}. This algorithm calculates a score for each data point based on the density of its local neighborhood compared to its neighbors' neighborhoods. Data points with a high LOF score are considered outliers.

\paragraph{Heuristic-based techniques} 
Heuristic-based techniques focus more on detection than correction. For instance, \cite{SurveyTrajectoryDM} does not replace outlier points with estimated values but instead removes them from the trajectory.
Common heuristics are based on speed with the idea that if the speed change rate is significantly higher than a given threshold and a proportion of the points in the entire trajectory, the point is removed. This approach has the advantage of not introducing any estimated values into the trajectory but can lead to significant data loss.

In \cite{movetk}, a new method category based on physical movement properties is introduced, such as speed (Optimal Speed-bounded) and acceleration (Optimal Acceleration-bounded). This method defines limits on the minimum and maximum allowed values for these properties and uses them to determine whether a point in the trajectory is consistent with the model. The limits are minimum v- and maximum speed v+, minimum a-, and maximum acceleration a+. It follows the definition that from one point to its successor, there should always be inside [v-, v+] and [a-, a+]. In addition, a trajectory T = ⟨p1,..., pn ⟩ is consistent with the model if and only if there exists at least one point in a path such that the measurement coincides with the point and the speed and acceleration are inside speed and acceleration bounds. The method defines a reachable region as a cone, i.e., given the physical boundaries, reaching the cone from point pi to pi+1 is possible. In contrast, it is not necessarily possible to construct a trajectory from the concatenation of two consistent sub-trajectories: the concatenation ⟨p1, ..., pn = q1, ..., qm⟩ of two consistent subsequences T = ⟨p1,..., pn ⟩ and U = ⟨q1,..., qm⟩ with pn = q1 is not necessarily consistent. Joining these sub-trajectories can reproduce inconsistent points—especially when considering an acceleration-bound model. The speed of two points can infer two accelerations for the same position. The model is called concatenable if it is possible to join both sub-trajectories, respecting the bounds.

In the next Section, we relate some of these methods to state-of-the-art libraries.

\section{State of Technology} 
\label{sec:stateoftheart}

This Section will review the available libraries that offer trajectory outlier detection and correction. We also relate some of the algorithms in Section \ref{sec:outlier} to state-of-the-art libraries.

MovingPandas \cite{movingPandas} \footnote{\url{https://github.com/anitagraser/movingpandas-examples}} is a Phyton library for trajectories of moving objects. Data can be represented in Pandas \cite{Pandas}, GeoPandas \cite{geopandas}, HoloViz \cite{holoviz}, CSV, GIS file formats, JSON, and geoJSON. MovingPandas implements structures for movement data in Python for interaction and analysis of movement. This library has many trajectory manipulation functions. Focusing on the outlier detection, this library implements KF. MovingPandas uses the Kalman Filter (KF) algorithm for outlier detection, which is implemented using the Stone Soup software \cite{stoneSoup}.

Scikit-mobility \cite{scikit}\footnote{\url{https://github.com/scikit-mobility/scikit-mobility}} is a Python library that extends Pandas \cite{Pandas}. Scikit-mobility offers functions for prepossessing and cleaning trajectory and analysis. The library chosen method for outlier detection is heuristic filtering, based on the speed and a given threshold. This approach can be effective for identifying points in the trajectory that deviate significantly from the expected pattern but may not be as accurate as other methods that use estimation to deal with outlier points.

Scikit-learn \cite{scikit-learn}\footnote{\url{https://scikit-learn.org/stable/}}  is a Python library focusing on the efficiency and ease of use of machine-learning algorithms. It provides simple and consistent interfaces, making it accessible to practitioners in machine learning. The library supports supervised and unsupervised learning algorithms standard linear regression models to neural networks, and it also includes tools for model fitting, data preprocessing, model selection, and evaluation. It applies the Local Outlier Factor (LOF) algorithm for outlier detection. The process begins by choosing features of interest that are then isolated for further processing. It then utilizes an unsupervised outlier detection method that computes the local density deviation of a data point with its neighbors. Outliers, regarded as samples having a considerably lower density than their neighbors, are determined by fitting the LOF model to the standardized data and identifying instances where the model's prediction is inaccurate.

Ptrail \cite{haidri2021ptrail}  \footnote{\url{https://github.com/YakshHaranwala/PTRAIL}} is a Python package that uses parallel computation and vectorization, making it suitable for large datasets. It offers several preprocessing steps, such as feature extraction, filtering, interpolation, and outlier detection. Ptrail removes outliers using a Hampel filter (HF) \cite{hampel} based on the distance and speed of the ships between consecutive points. For each trajectory, the method calculates the median of a sliding window and adjacent points on each side of the trajectory. The HF also estimates the standard deviation of each point about its window median using the median absolute deviation. If a measurement differs from the median by more than the threshold multiplied by the standard deviation, the filter replaces the sample with the median.

PyMove \cite{Pymove1,Pymove2}\footnote{\url{https://github.com/InsightLab/PyMove}}  is a Python library. It offers a range of operations for data preprocessing and pattern mining. PyMove also provides tools for data visualization, allowing users to explore and understand their data through various techniques and channels. PyMove detects outlier points considering the distance traveled, minimum and maximum speed.

Movetk \cite{movetk}\footnote{\url{https://github.com/movetk/movetk}} is C++ library. It offers tools for constructing, cleaning, and analyzing trajectory data. One of the key features of Movetk is its implementation of the Optimal Speed-bounded and Optimal Acceleration-bounded algorithms for outlier detection. Movetk implements outlier detection methods based on the Optimal Speed-bounded algorithm and the Optimal Acceleration-bounded algorithm. In addition to these algorithms, Movetk implements various other methods for outlier detection, including greedy and smart greedy approaches. These methods build on the basic speed and acceleration-bounded algorithms and incorporate additional strategies and techniques to improve their performance and accuracy. For the greedy approach, Movetk greedily builds a consistent subsequence by testing if the new measurement is consistent with the last in the subsequence under the speed-bounded model (GSB) or acceleration-bounded model (GAB). In addition, the authors implement a Smart Greedy Speed/Acceleration-bounded method (SGSB/SGAB). With SGSB and SGAB, multiple subsequences are tracked simultaneously. The next measurement is added to all subsequences that end in a consistent measurement; if there is no such subsequence, a new subsequence starting with the measurement is created. In the end, the longest subsequence is returned. Also, as a baseline, the library implements their interpretation of the method described in \cite{SurveyTrajectoryDM} as Local Greedy Speed-bounded (LGSB). In LGSB, a graph is constructed with a vertex per measurement. Two vertices are connected if their timestamps are successive in the original trajectory and they are consistent with the speed-bound. A measurement is added to the output if and only if its vertex is in a connected component of a given size. LGSB does not guarantee that the complete output is consistent according to the speed bound \cite{SurveyTrajectoryDM}.

MEOS \footnote{\url{https://libmeos.org}, \url{https://github.com/MobilityDB/}} is an open-source C library engineered for mobility analytics built upon \cite{MobilityDB}. It is a highly versatile solution, providing a single source code that integrates with multiple programming languages. MEOS's flexibility allows it to operate across diverse computing environments, ranging from edge to cloud-based platforms, and efficiently manage batch and stream processing. The library detects outliers using a heuristic method comparing maximum and minimum speeds.

Argosfilter \cite{argosfilter} \footnote{\url{https://cran.r-project.org/web/packages/argosfilter/argosfilter.pdf}}  is an R package that offers a set of functions for working with trajectory data. The outlier detection in Argosfilter uses two different methods: one based on speed and the other based on location. The speed-based method is similar to the one provided in \cite{SurveyTrajectoryDM}, but the location-based method is based on the algorithm described in the paper \cite{argosfilter}. This method uses a set of spatial constraints, such as a minimum distance between two points or a maximum distance from a reference point, to identify and remove outliers from a trajectory.

Stmove \cite{stmove}\footnote{\url{https://tinyurl.com/stmove}} is an R package library. It provides construction functions, filter, and outlier detection functions. For the outlier filter, the KF is applied.

MVOutlier \cite{mvoutlier} \footnote{\url{https://cran.r-project.org/web/packages/mvoutlier/index.html}}  is R package is a tool for identifying outliers in multivariate data. One of the primary strategies employed by this package is distance-based outlier detection, in which the distance of each observation from others or a central direction measure is calculated. Observations falling beyond a predefined distance threshold are then tagged as outliers. Among the methods provided by the MVOutlier package include UniOD, which carries out univariate outlier detection for each variable individually, and Doutlier, which performs depth-based detection of multivariate outliers. The package also incorporates the Minimum Covariance Determinant (MCD) method and the Mahalanobis distances method, which is adept at accommodating the covariance between variables.  Furthermore, the Stahel-Donoho estimates present another robust method for location and scatter estimates, which is beneficial for outlier detection.

\begin{table}[]
\resizebox{\textwidth}{!}{%
\begin{tabular}{|c|c|l|c|}
\hline
\textbf{Name} & \textbf{Outlier Detection Method} & \textbf{Category} & \textbf{Language} \\ \hline
Ptrail & Hampel identifier & Statistic-based & Python \\ \hline
MovingPandas & Kalman filter & Statistic-based & Python \\ \hline
Scikit Mobility & Maximum speed & Heuristic-based & Python \\ \hline
Pymove & Maximum/minimal speed & Heuristic-based & Python \\ \hline
Scikit-learn & Local Outlier Factor & Graph-based & Python \\ \hline
Stmove & Kalman & Statistic-based & R \\ \hline
Argosfilter & Maximum/minimal speed & Heuristic-based & R \\ \hline
MVOutlier & Minimum Covariance Determinant & Graph-based & R \\ \hline
MoveTk & Linear speed bound & Heuristic-based & C++ \\ \hline
MEOS & Maximum/minimal speed & Heuristic-based & C \\ \hline
\end{tabular}%
}

\label{tab:stateOftechnology}
\caption{Name, method for outlier detection, method's category, and programming language used by each considered library}
\end{table}

In the next Section, we present experiments performed in the libraries presented above.

\section{Experiments}
\label{sec:experiments}

This Section presents our method for constructing ground-truth in Subsection \ref{subec:gorundThuth}. We present an analysis of the experiment results in Subsection \ref{subec:study}.

\subsection{Constructing Ground-truth}
\label{subec:gorundThuth}

Constructing a ground-truth for trajectory data analysis is a task that presents several challenges. The difficulties include the cost, accuracy, and scale of the required ground-truth data. Additionally, cleaning the data to remove errors and noise can be time-consuming and may only sometimes produce reliable results.

{\color{blue}We present a method that cross-references data that originate from multiple sensors that belong to the same source. For example, the moving object has multiple sensors, such as GPS, gyrocompass, and speed sensors. Hence, the receiver receives multiple signals for the same source. Our method constructs the ground-truth by cross-referencing these signals. The intuition is that it is possible to receive faulty data for one sensor, but it much less the received data of the other sensors will also be faulty at the same time.} 

\begin{figure}[!htb]
	\includegraphics[width=0.70\textwidth]{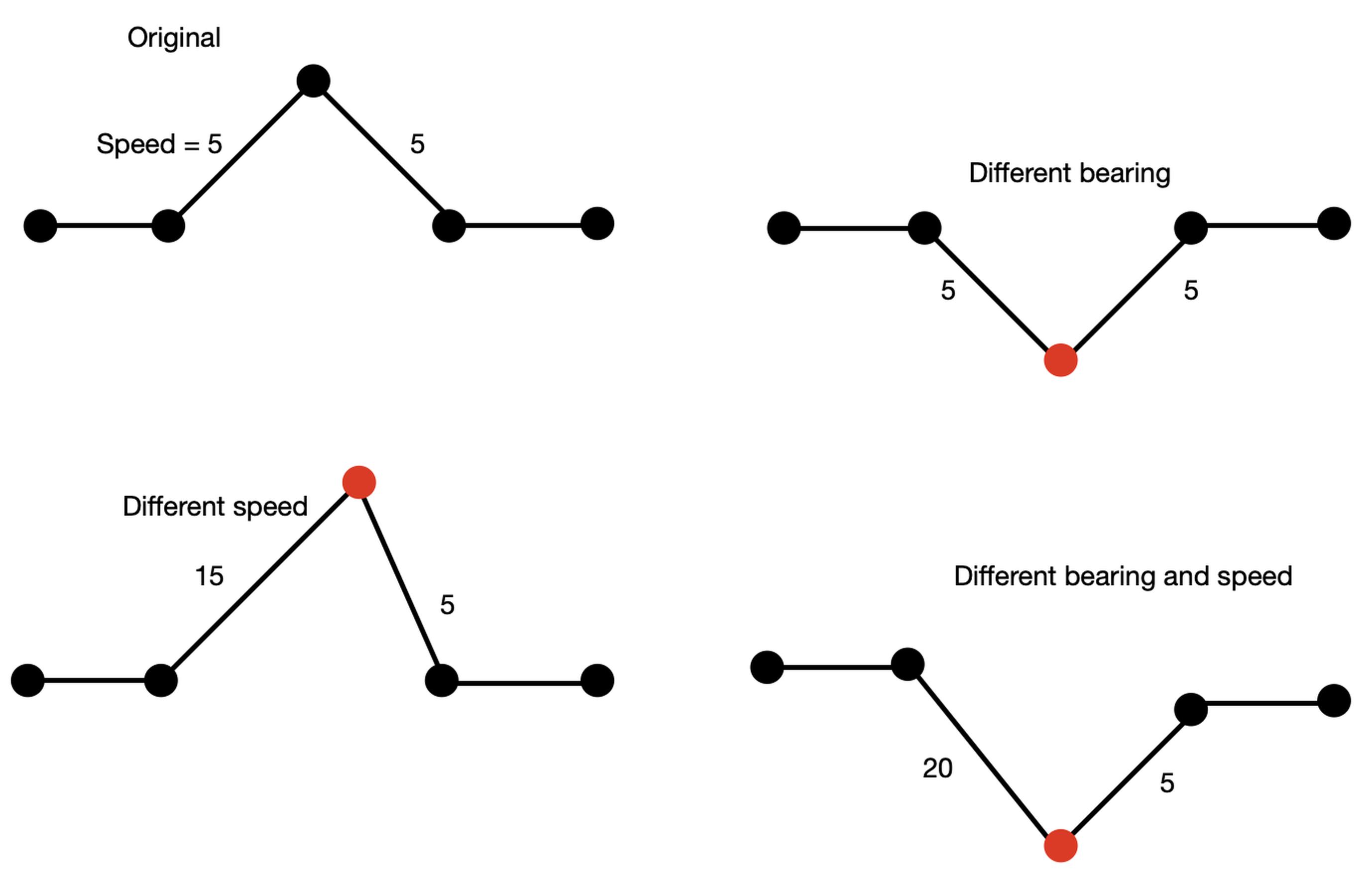}
	\caption{The ground-truth method can detect variation in speed and/or bearing from the original trajectory}
	\label{fig:Differentcases}
\end{figure}

Hence, our proposed method offers a different approach to traditional methods for constructing ground-truth, such as manual annotation or data cleaning techniques like those described in \cite{song,datacleaning,anotation}. It also allows for data integration from multiple sensors, increasing the overall accuracy of the ground-truth. We consider speed and bearing to accommodate both errors in direction and speed. Figure \ref{fig:Differentcases}
illustrates the different errors one can encounter in a trajectory for speed and bearing. The Algorithm calculates and compares the speed and bearing with the recorded values in the data. The input consists of speed and heading thresholds (tS and tH) and the file path. In lines 6--12, for each point in the file, firstly, we check if the IDs are the same, i.e., the points belong to the same trajectory. Secondly, we calculate the speed and bearing between the point and its successor. Later, we compare the newly calculated speed and bearing with the files' input in line \ref{alg:iffspeed}. If the difference in speed or heading is bigger than the inputted thresholds, the point is considered an outlier and added to the output array. {\color{blue}The proposed method involves iterations to define an optimal threshold for the studied parameters: heading and speed. We have used visualizations of original trajectories and corrected trajectories to validate the ground-truth method. The fine-tuning can be validated by splitting the dataset and cross-validating visually. In the future, we plan to develop a self-tuning algorithm for thresholds based on machine learning. In addition, creating an artificial dataset where we can add errors/outliers to validate the method better.}

\begin{algorithm}[h!]
\scriptsize
\KwIn{thresoldSpeed $tS$, thresoldHeading $tH$, FilePath $file$}
\KwOut{Outliers}
Outliers $ \leftarrow \{\}$\;
\texttt{file} $\leftarrow$ \textsc{ReadFile} (\texttt{FilePath})\;
\texttt{file} $\leftarrow$ \textsc{OrderByIDandTimeStamp} (\texttt{file})\;
$speed \leftarrow 0$\;
$bearing \leftarrow 0$\;
\ForEach{$point \in \texttt{file.length}$}{
    \If{\texttt{file[point].id == file[point+1].id}}{
        $speed \leftarrow \textsc{CalculateSpeed} (\texttt{point}, \texttt{point}+1)$\;
        $bearing \leftarrow \textsc{CalculateBearing} (\texttt{point}, \texttt{point}+1)$\;
        \If{ $\left| speed - \texttt{file[point].sensorRecordedSpeed} \right| \geq tS$ \textbf{OR} \\
             $\left| bearing - \texttt{file[point].sensorRecordedBearing} \right| \geq tH$}{
            $\texttt{Outliers.append(point)}$\;
        }
    }
}
\Return{Outliers}
\caption{Ground-truth}
\label{alg:iffspeed}
\end{algorithm}

{\color{blue}Our method involves cross-referencing the data from multiple sensors to detect discrepancies and annotate them as ground-truth for outliers .} We apply this method to data from modern multi-sensor tracking, such as GPS, AIS, ADS-B, Mode S, TCAS, and FLARM sensors. By combining the data from multiple sensors, we can increase the overall accuracy of the ground-truth in the presence of errors or noise in individual sensors. When comparing data from different sensors, we can analyze the data to see if there are any discrepancies or inconsistencies between all sensors. In addition, statistical tests can determine whether the data is consistent with a particular hypothesis or model. For example, we can check the calculated speed with the speed given in the data.

{\color{blue}Our datasets for this experimental study consist of multi-sensor trajectory data. Each dataset contains data from different sensors. For example, AIS data is composed of data provenient from multiple sensors, such as the speed captured by the sensor and location informed by the GPS.} The first sensors is AIS data: {\color{blue} 
AIS is the location tracking system for sea vessels. This data is collected from participating ships worldwide, providing diverse trajectories. AIS data is composed of data provenient from multiple sensors. An AIS message is composed of 26 elements. These messages contain information from sensors such as Gyrocompass, Rate of Turn Indicator, GPS, VHF radio frequency transceivers, Information Systems (ECDIS), and radar. Each data point provides information on timestamp, type of mobile, id, latitude, longitude, heading, speed over ground, course over ground, ship type, cargo type, width, length, and destination, among other information. In this paper, we utilize a total of 4.3 GB.
}

\begin{figure}[h]
    \centering
    \begin{minipage}[b]{0.45\textwidth}
        \includegraphics[width=\textwidth]{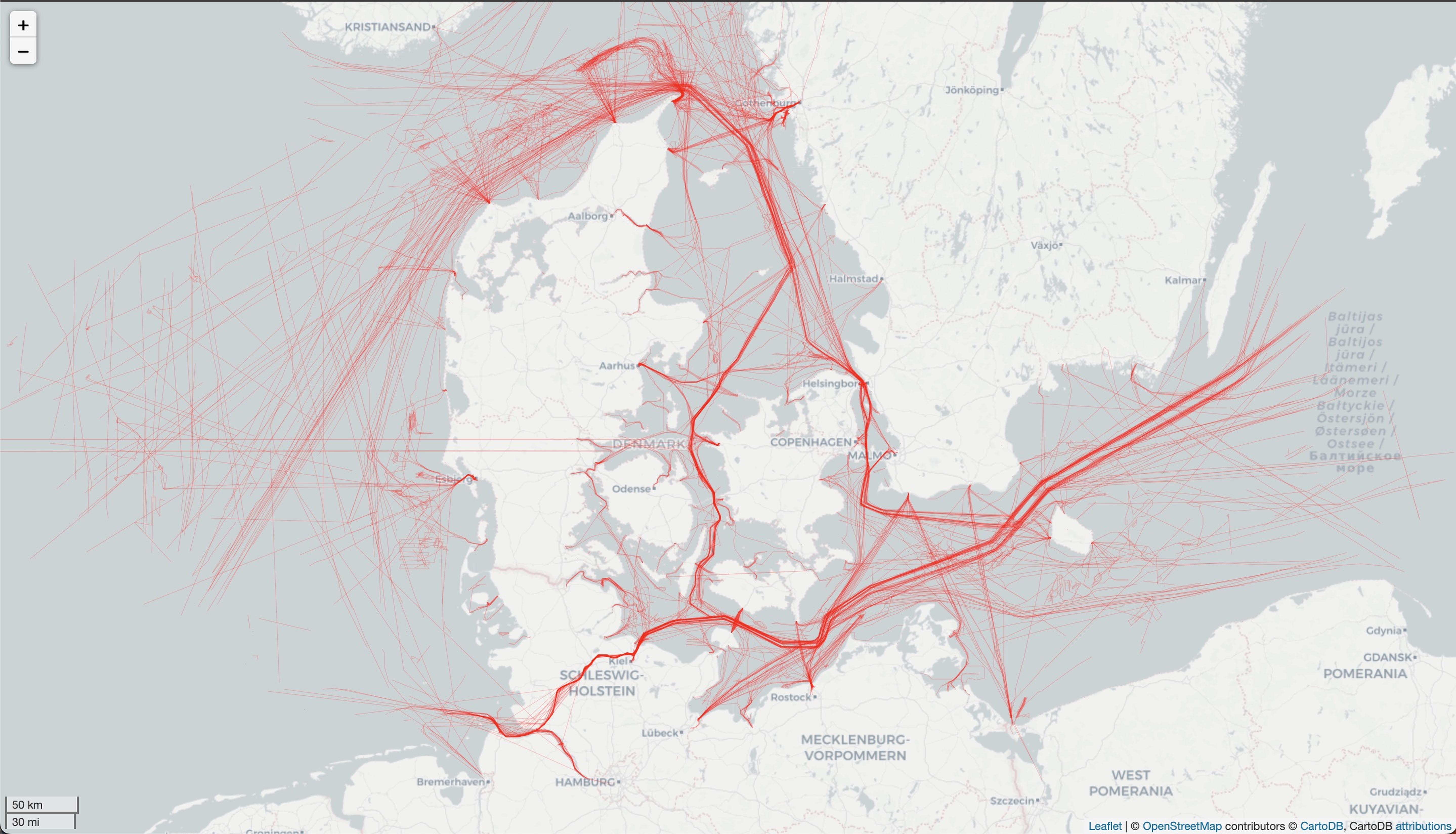}
        \caption{AIS raw dataset}
        \label{fig:ais}
    \end{minipage}
    \hfill
    \begin{minipage}[b]{0.45\textwidth}
        \includegraphics[width=\textwidth]{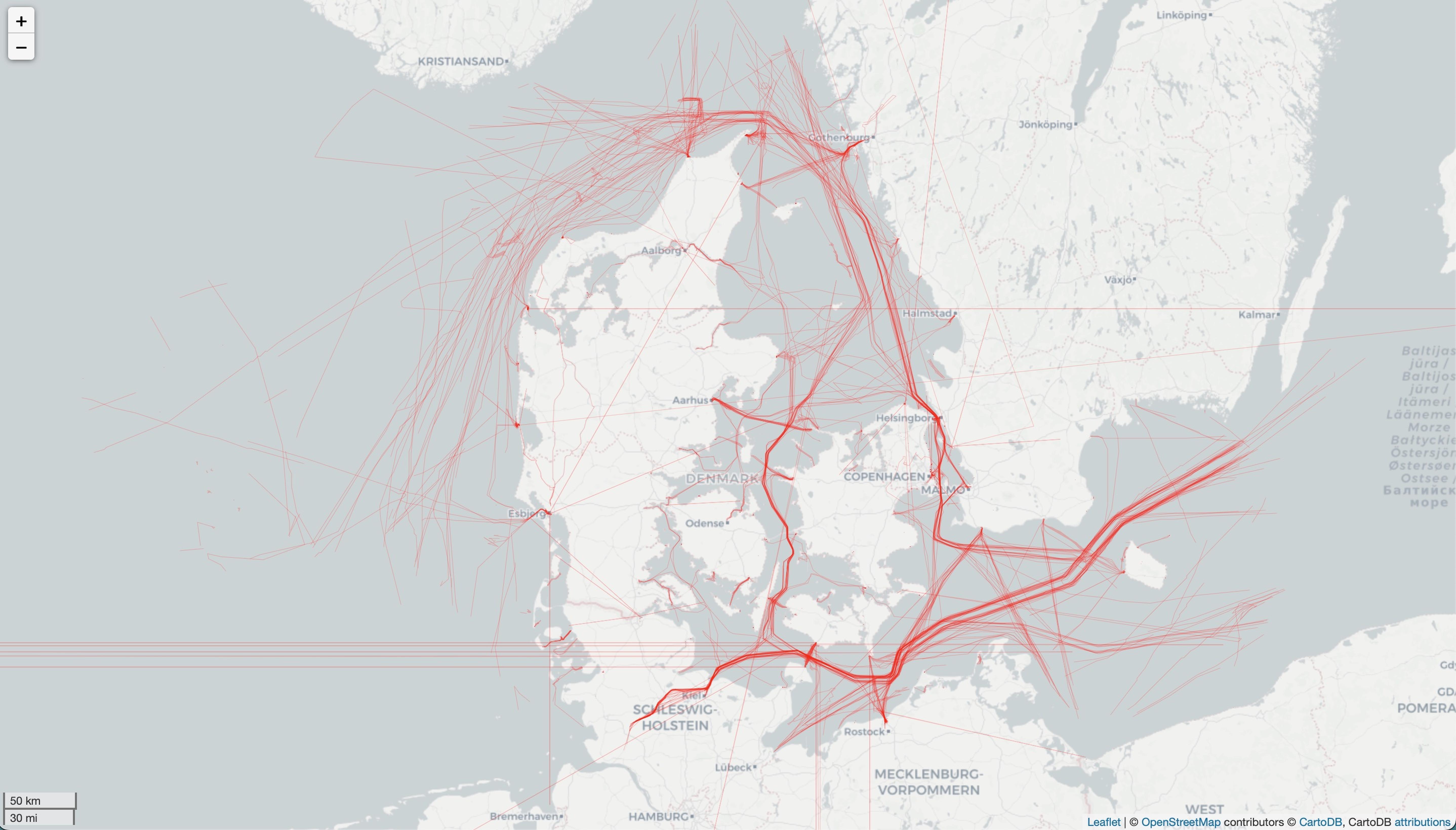}
        \caption{AIS with ground-truth filtering}
        \label{fig:AISnormal}
    \end{minipage}
\end{figure}

{\color{blue} OpenSky Network\footnote{\url{https://opensky-network.org}} offers raw data stored in a historical database to study and improve air traffic control technologies and processes. Fig.4 shows an illustration of the raw data. Each data point in the dataset has time, id, latitude, longitude, speed, heading, Barometer altitude and altitude, timestamp, and other additional information.}

\begin{figure}[h]
    \centering
    \begin{minipage}[b]{0.45\textwidth}
        \includegraphics[width=\textwidth]{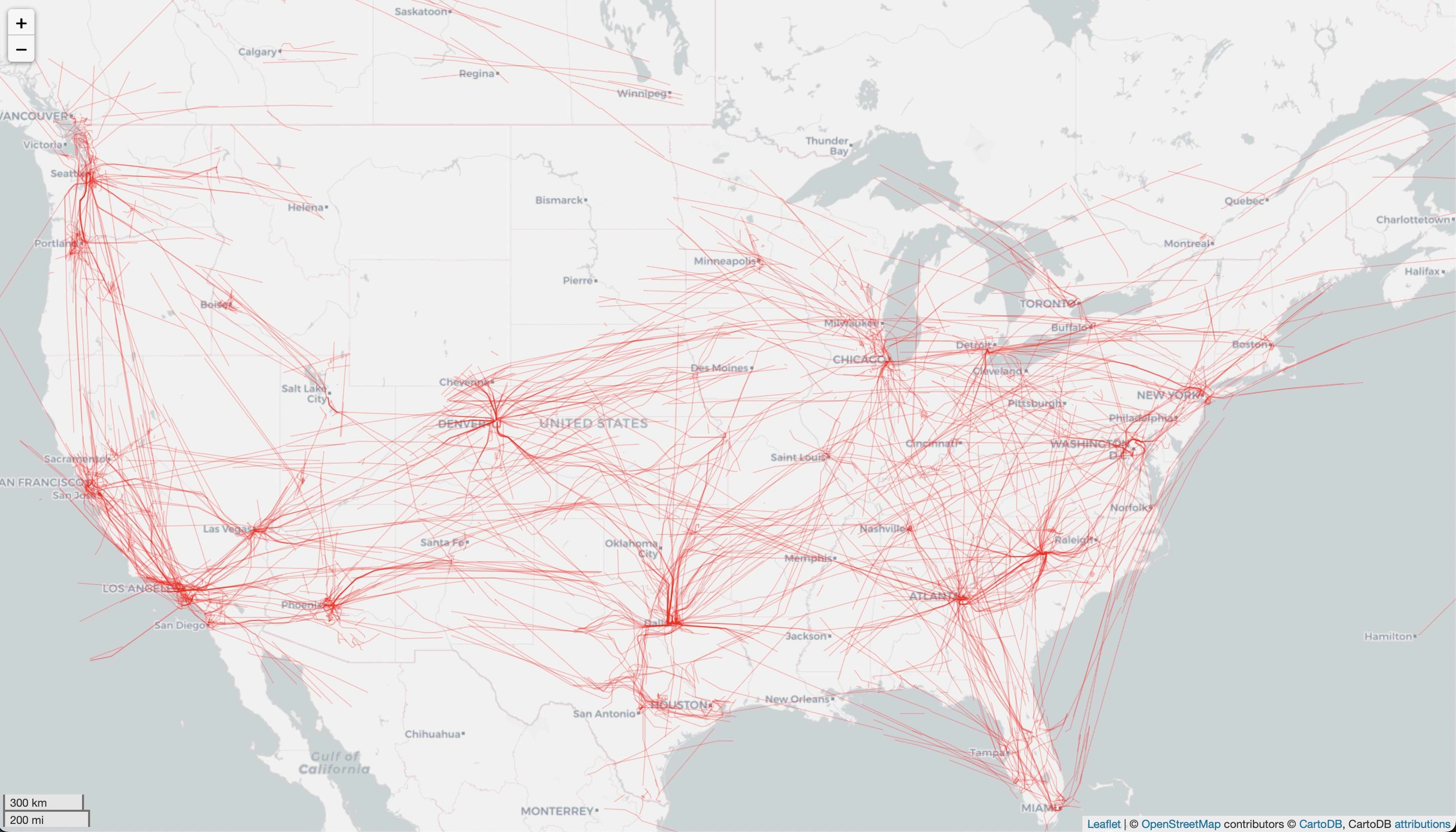}
        \caption{Open sky raw dataset}
        \label{fig:OpenSky}
    \end{minipage}
    \hfill
    \begin{minipage}[b]{0.45\textwidth}
        \includegraphics[width=\textwidth]{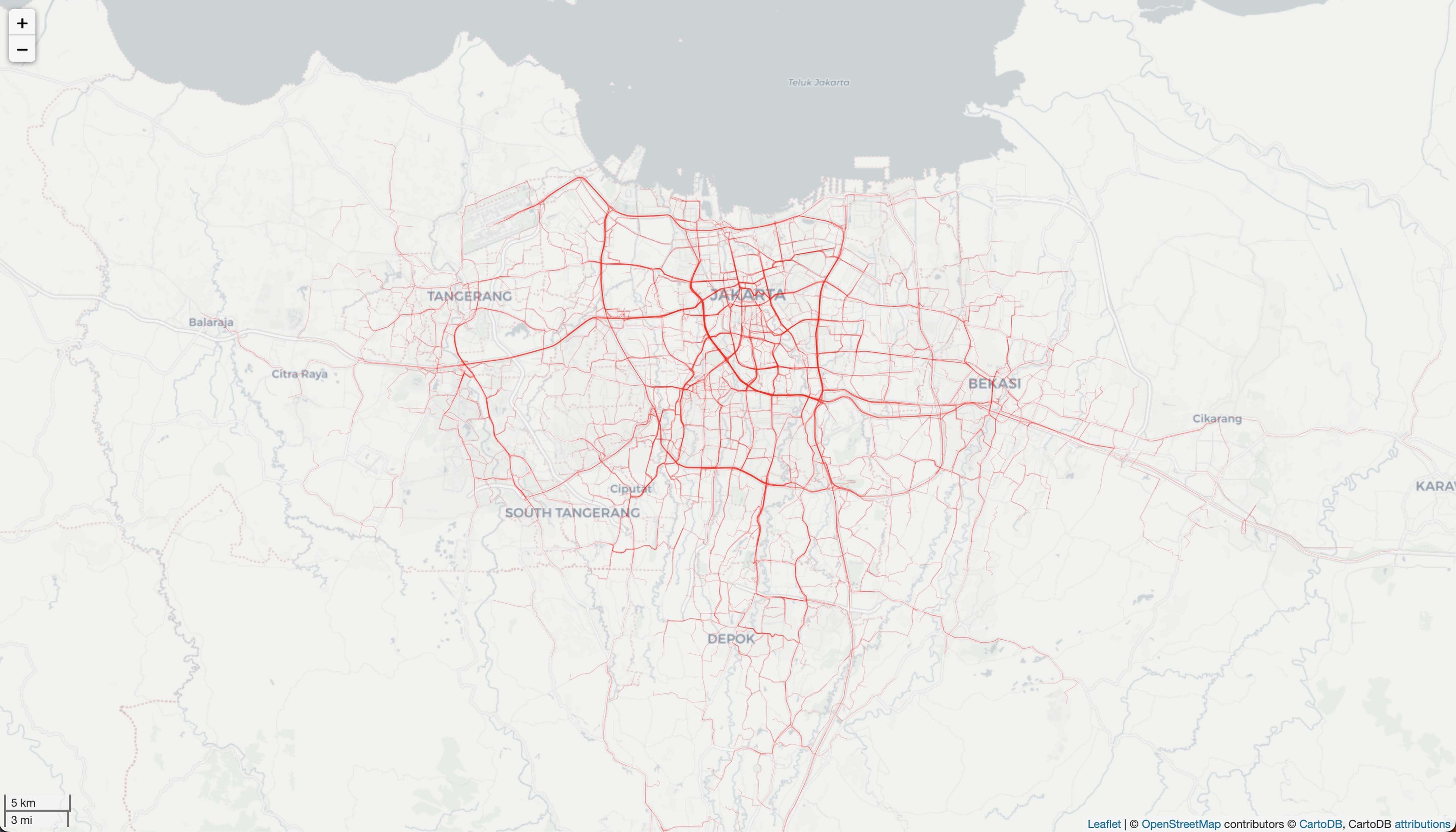}
        \caption{Jakarta raw dataset}
        \label{fig:jackarta}
    \end{minipage}
\end{figure}

\begin{figure}[!htb]
	\includegraphics[width=0.40\textwidth]{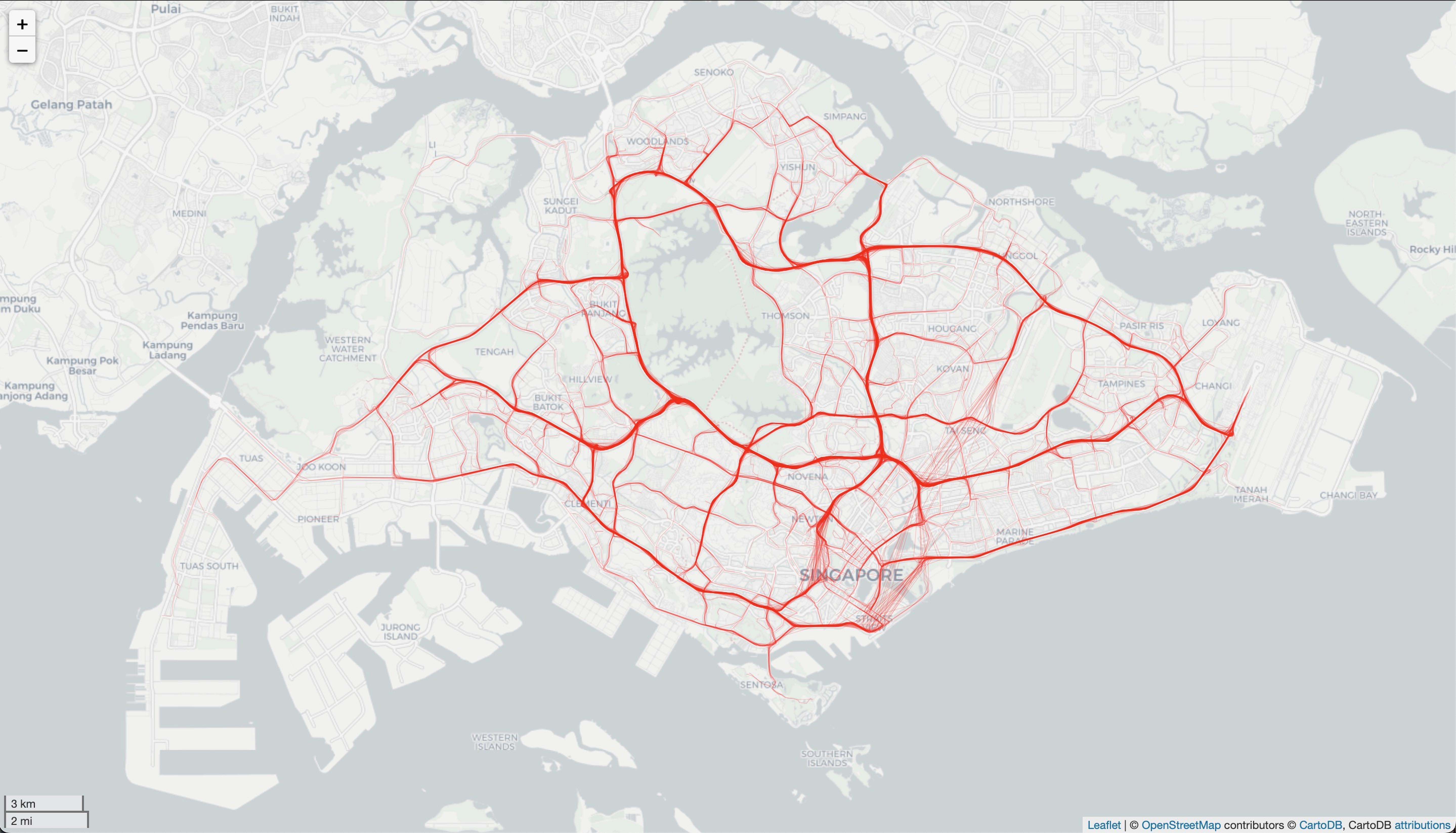}
	\caption{Singapore raw dataset}
	\label{fig:Singapore}
\end{figure}

{\color{blue}Jakarta and Singapore datasets consist of GPS trajectories from Grab Tech \footnote{\url{https://www.grab.com/sg/}} drivers’ phones in Singapore and Jakarta cities in southeast Asia. Grab-Posisi \cite{citiesData} covers over 1 million kilometers and over 88 million points. Each trajectory has information on speed, heading direction, area, and distance covered during the travel and duration.} Images \ref{fig:jackarta} and \ref{fig:Singapore} show the raw data.

\begin{table}[]
\resizebox{\textwidth}{!}{%
\begin{tabular}{|c|l|c|c|c|c|}
\hline
\textbf{DataSet} & \textbf{Description} & \textbf{Source} & \textbf{Total Points} & \textbf{Outliers} & \% \\ \hline
Jakarta & Vehcile GPS  trajectories& Grab-Posisi dataset \cite{citiesData}& 16798804 & 85275 & 1.523\% \\ \hline
Singapore & Vehicle GPS  trajectories& Grab-Posisi dataset \cite{citiesData} & 9096633 & 68202 & 2.249\% \\ \hline
OpenSky & \begin{tabular}[c]{@{}l@{}}ADS-B, Mode S, \\ TCAS and FLARM\end{tabular} & OpenSky Network \footnote{\url{https://opensky-network.org}} &5417090 & 2 & 0.000\% \\ \hline
AIS & Ship tracks & Danish Maritime Authority \footnote{\url{https://dma.dk/}}&25714294 & 957 & 0.011\% \\ \hline
\end{tabular}%
}
\label{table:datasetOutliers2}
\caption{Datasets total points, data source, outliers, correct points, and percentage of outlier over total points}
\end{table}

% description and source (grab dataset, remove normal points collumn)

Table 2 shows the total number of records, the number of outliers detected in the cross-check, the correct points, i.e., points not considered outliers, and the percentage of outliers over total points. The cross-check method was used to detect the outliers.

OpenSky data does not have a significant amount of outliers.
One possible reason for the low number of outliers in the OpenSky data could be the accuracy of the sensors used to collect the data. If the sensors are highly accurate, there may be fewer errors or discrepancies in the data, resulting in more occasional outliers. Additionally, the data may have been adjusted or corrected in some way to account for any errors or noise, further reducing the number of outliers. 

%{\color{blue}To evaluate the precision of various outlier detection libraries, we employ datasets such as AIS and OpenSky, which are characterized by minimal outliers. This ensures that the libraries are not overly aggressive in data point removal. In contrast, we utilize datasets from Jakarta and Singapore for assessing recall, which presents a higher rate of outliers.}

\subsection{Experimental Study}
\label{subec:study}

{\color{blue}The experimental study proposed in this paper evaluates the libraries, and it consists of the compilation of various mobility datasets originating from multiple sensors, the compilation of different evaluation metrics, and the generation of ground-truth for each of these datasets. Furthermore, several libraries are assessed through this experiment. The metrics to be measured are recall, precision, F-2, and F-0.5.}

\begin{figure}
  \centering
  % include first image
  \includegraphics[width=1.0\linewidth]{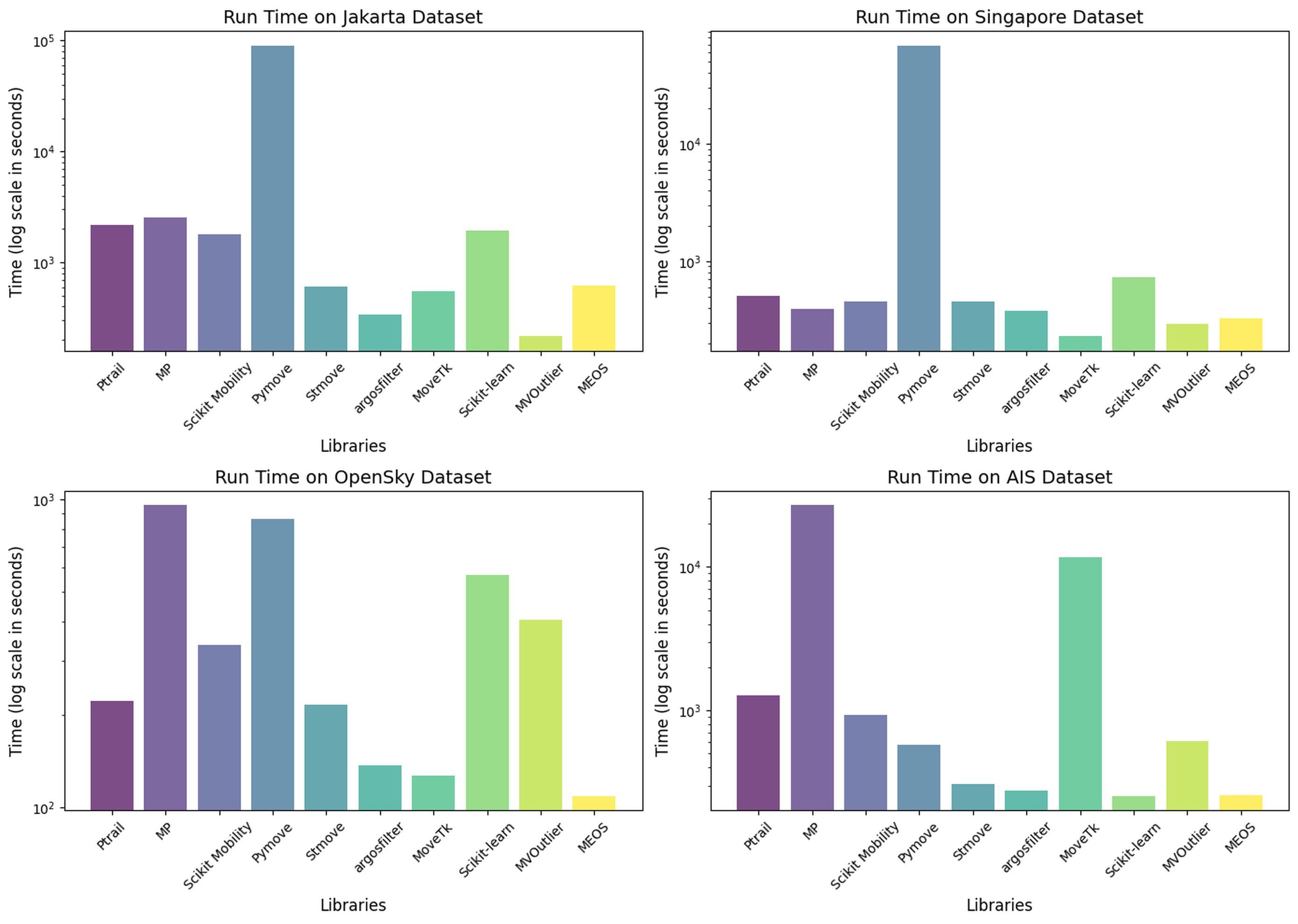}  
  \caption{Run time for libraries Ptrail, MovingPandas, Scikit-mobility, PyMove, Stmove, Argosfilter, MoveTK, Scikit-learn, MVOutlier and MEOS. The top left graph corresponds to Jakarta datasets, top right graph represents the Singapore dataset, the bottom left graph  represents the OpenSky dataset run time and the bottom right is AIS dataset run time}
  \label{fig:RunTime}
\end{figure}

  The experiments were run in ten libraries composed of MovingPandas, Scikit-mobility, Scikit-learn, Ptrail, PyMove, movetk, MEOS, Argosfilter, Stmove, and MOutlier. These were described in Section \ref{sec:stateoftheart}. Fig.~\ref{fig:RunTime}. The data used in the experiment was described in Section \ref{subec:gorundThuth}. The source code is publicly available. \footnote{\url{
https://github.com/marianaGarcez/OutlierDetectionLibraries}}

 The European aviation industry \cite{Europeanflights}\footnote{\url{https://www.eurocontrol.int/publication/objective-skygreen-2022-2030}} developed methods to reduce carbon emissions to meet climate targets. An aircraft can fly an optimal flight path and use various technologies and infrastructure to minimize fuel consumption and carbon emissions. This might include using modern flight planning software and meteorological data to plan for minimal fuel, using green energy at the airport to power the aircraft on the ground, and using electric taxi solutions to minimize ground-based emissions. The aircraft would also fly an optimal climb phase, follow a fuel-efficient cruising level, and use idle thrust descent to reduce fuel consumption during descent, i.e., the aircraft should change its altitude as rarely as possible. Due to this standard, we consider the OpenSky dataset to have only 2D rather than 3D data. However, it is essential to note that the assumption may not hold true in all circumstances. It may be necessary to use 3D data or incorporate altitude data in some cases to analyze flight patterns accurately \cite{attia2009spatiotemporal}.

In Figure~\ref{fig:RunTime}, a comparative analysis of the run times for various libraries is presented across four distinct datasets: Jakarta, Singapore, OpenSky, and AIS. Notably, the Pymove library consistently exhibits the highest run times across all datasets, with an increase in computational time for larger datasets such as Jakarta and Singapore.

Libraries developed in R programming language offer comparable run times. Similarly, libraries based on C++ (MoveTk) and C (MEOS) demonstrate run times similar to those of the R libraries. Furthermore, Scikit-learn exhibits competitive performance, likely attributed to its highly optimized Python implementation and algorithm. The MoveTk library displays an anomalous behavior in the execution time with the AIS dataset. This irregularity could be attributed to the underlying process employed for data reading. Given these observations, careful consideration of run times is advised when selecting a library for trajectory data analysis and modification.

\begin{figure}[h]
    \centering
    \begin{minipage}[b]{0.49\textwidth}
        \includegraphics[width=\textwidth]{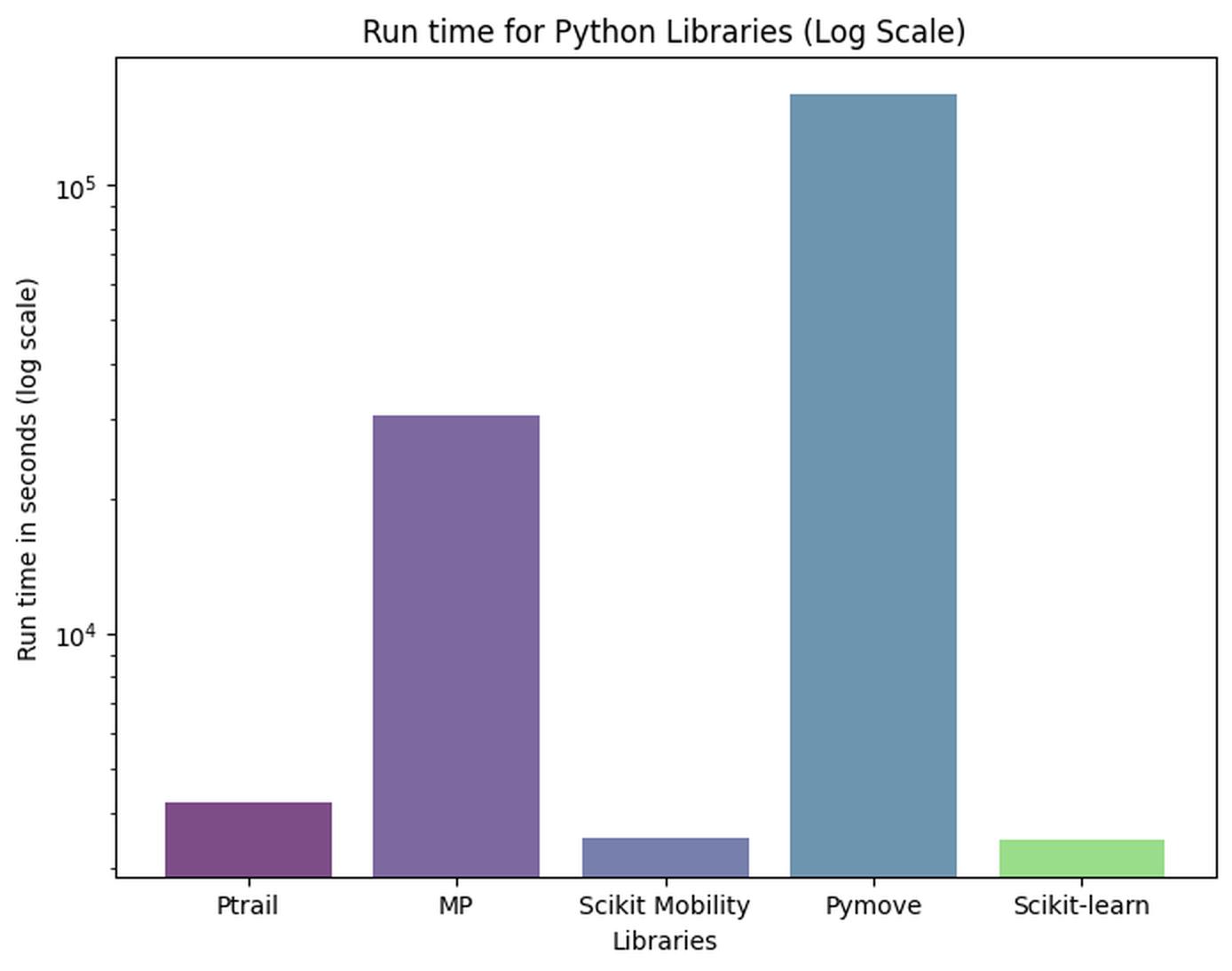}
        \caption{Sum of run time across all datasets for Python libraries Ptrail, MovingPandas,\\ Scikit-mobility, PyMove, and Scikit-learn}
        \label{fig:RunTimeP}
    \end{minipage}
    \hfill
    \begin{minipage}[b]{0.49\textwidth}
        \includegraphics[width=\textwidth]{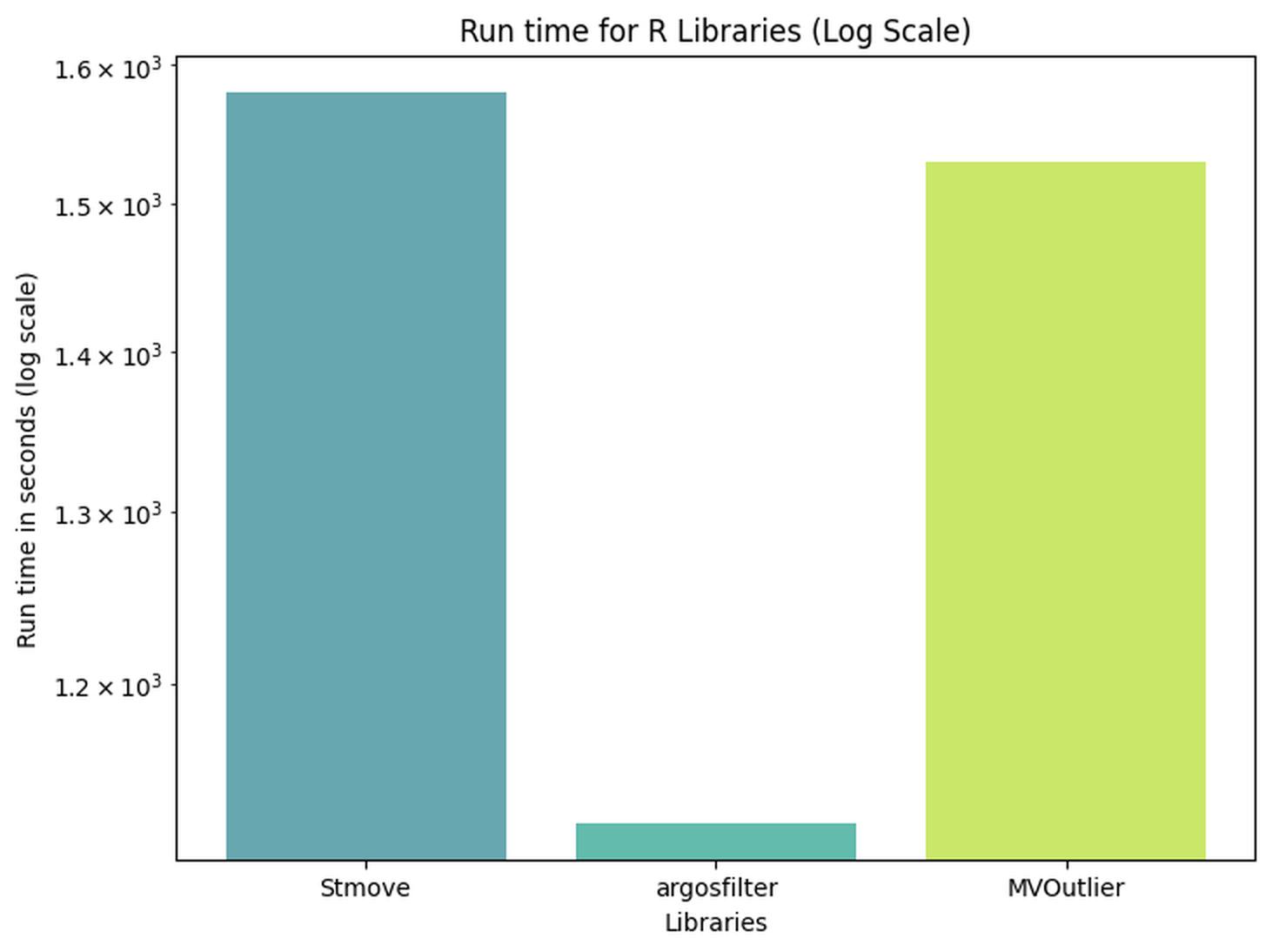}
        \caption{Sum of run time across all datasets for R libraries Stmove, Argosfilter, and MVOutlier}
        \label{fig:RunTimeR}
    \end{minipage}
\end{figure}

In Figures~\ref{fig:RunTimeP} and ~\ref{fig:RunTimeR}, we see the total run times across libraries developed in Python and R, respectively. What stands out is an association between the language of implementation and the efficiency of the library in question. To summarize, the choice of programming language appears to substantially impact the run time, emphasizing the need to consider language compatibility and efficiency when evaluating libraries for specific tasks.

\begin{figure}[h!]
  \centering
  % include first image
  \textbf{ \centering
  % include second image
  \includegraphics[width=0.8\linewidth]{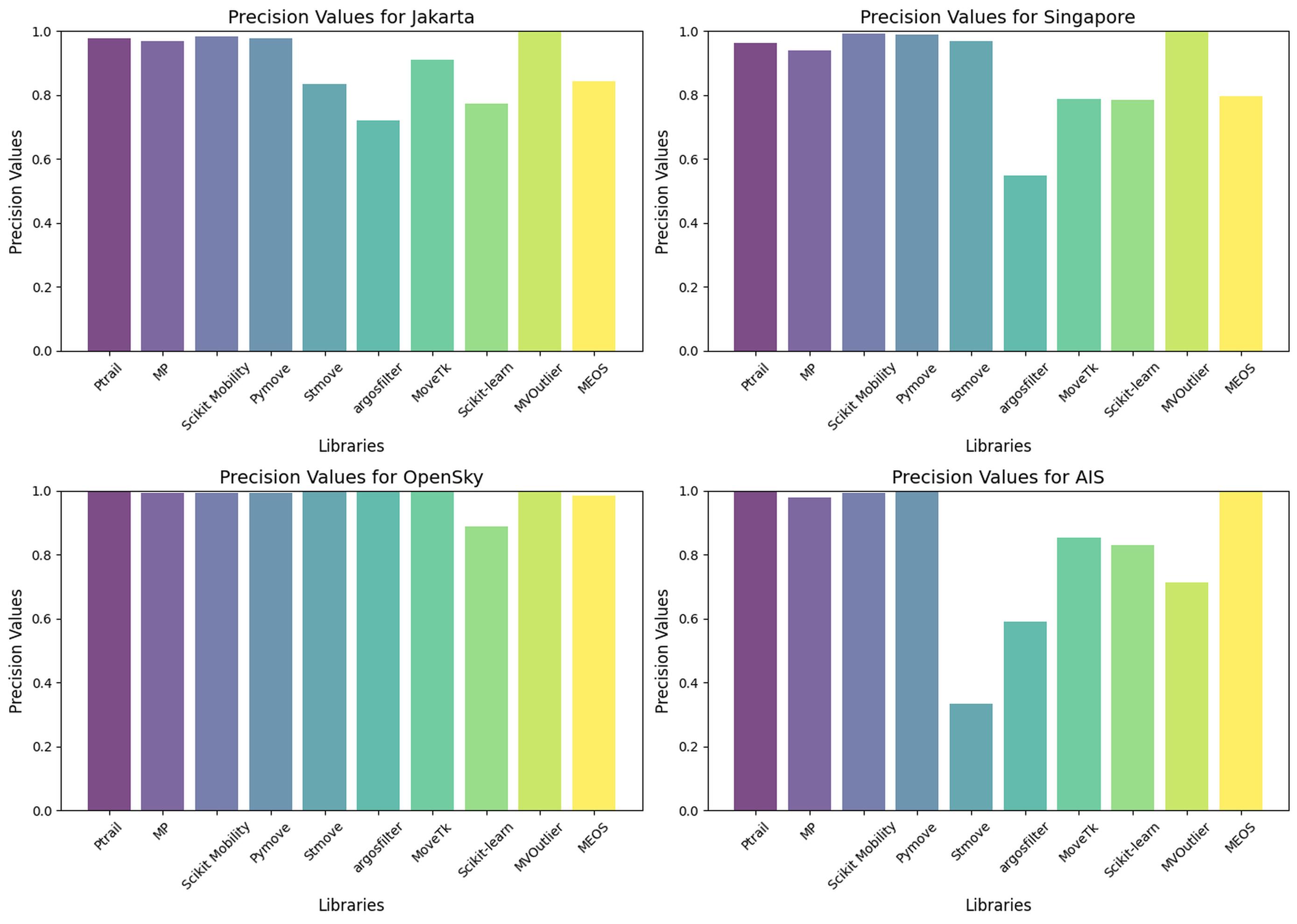}  
  \caption{Precision score for libraries Ptrail, MovingPandas, Scikit-mobility, PyMove, Stmove, Argosfilter, MoveTK, Scikit-learn, MVOutlier and MEOS. The top left graph corresponds to Jakarta datasets, the top right graph represents the Singapore dataset, the bottom left graph  represents the OpenSky dataset and the bottom right is the AIS dataset}
  \label{fig:Precision}}
  \includegraphics[width=0.8\linewidth]{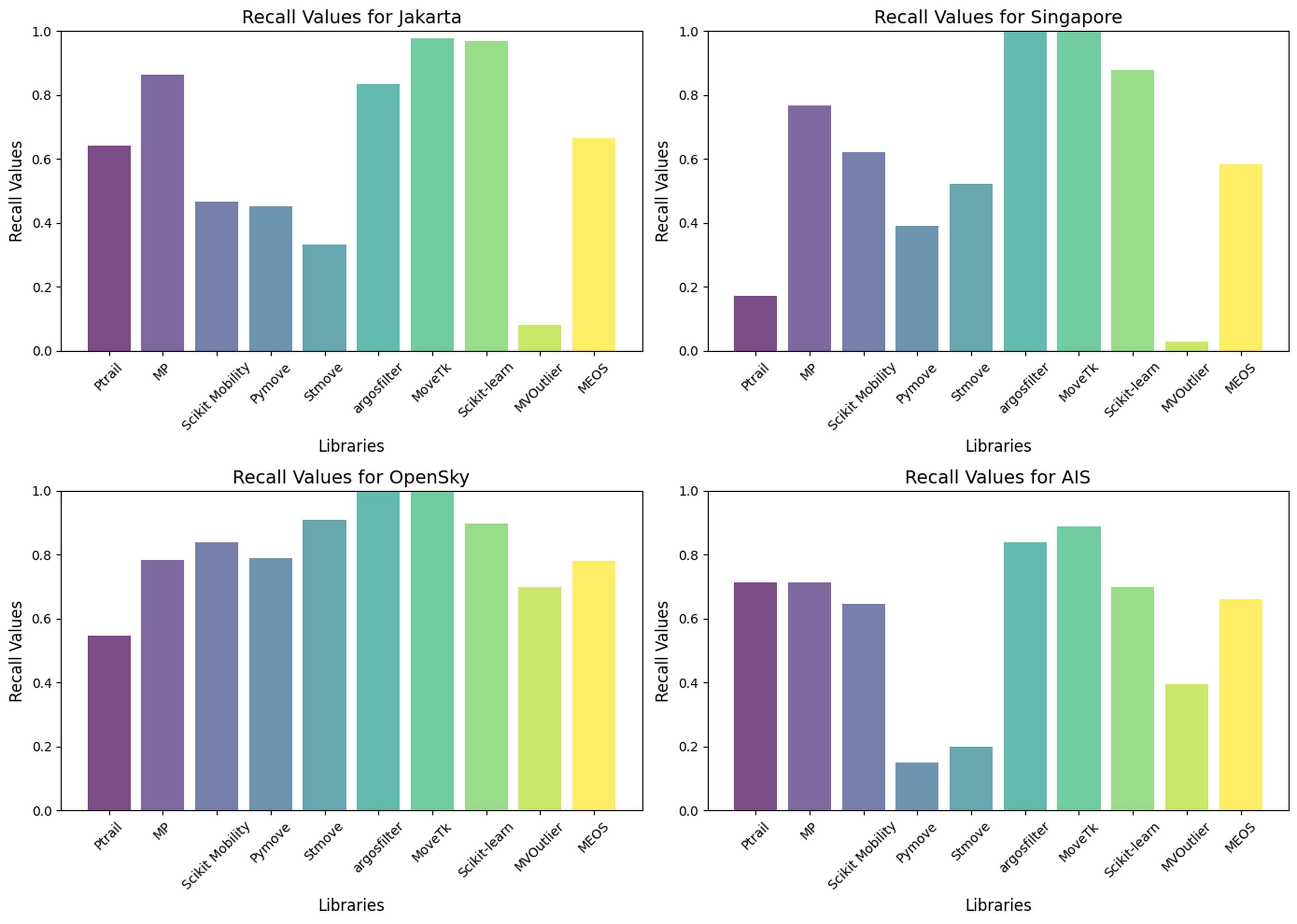}  
  \caption{Recall for libraries Ptrail, MovingPandas, Scikit-mobility, PyMove, Stmove, Argosfilter, MoveTK, Scikit-learn, MVOutlier and MEOS. The top left graph corresponds to Jakarta datasets, the top right graph represents the Singapore dataset, the bottom left graph  represents the OpenSky dataset and the bottom right is the AIS dataset}
  \label{fig:LibrairesRecall}
\end{figure}

\begin{figure}
  \centering
  \includegraphics[width=0.85\linewidth]{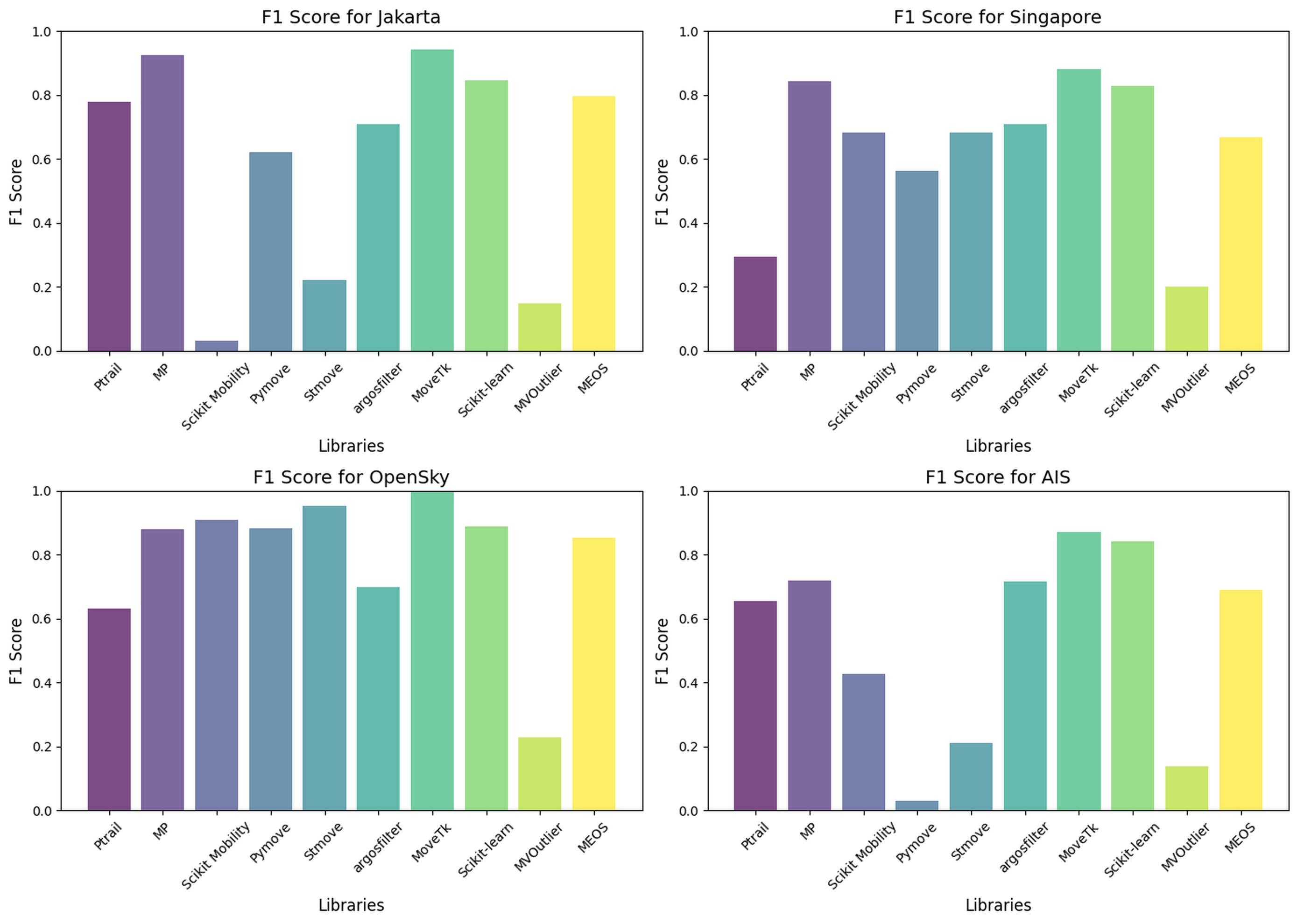}  
  \caption{F-1 score for libraries Ptrail, MovingPandas, Scikit-mobility, PyMove, Stmove, Argosfilter, MoveTK, Scikit-learn, MVOutlier and MEOS. The top left graph corresponds to Jakarta datasets, the top right graph represents the Singapore dataset, the bottom left graph represents OpenSky dataset and the bottom right is the AIS dataset}
  \label{fig:LibrairesF1}
\end{figure}

\begin{figure}[h!]
  \centering
  \includegraphics[width=0.85\linewidth]{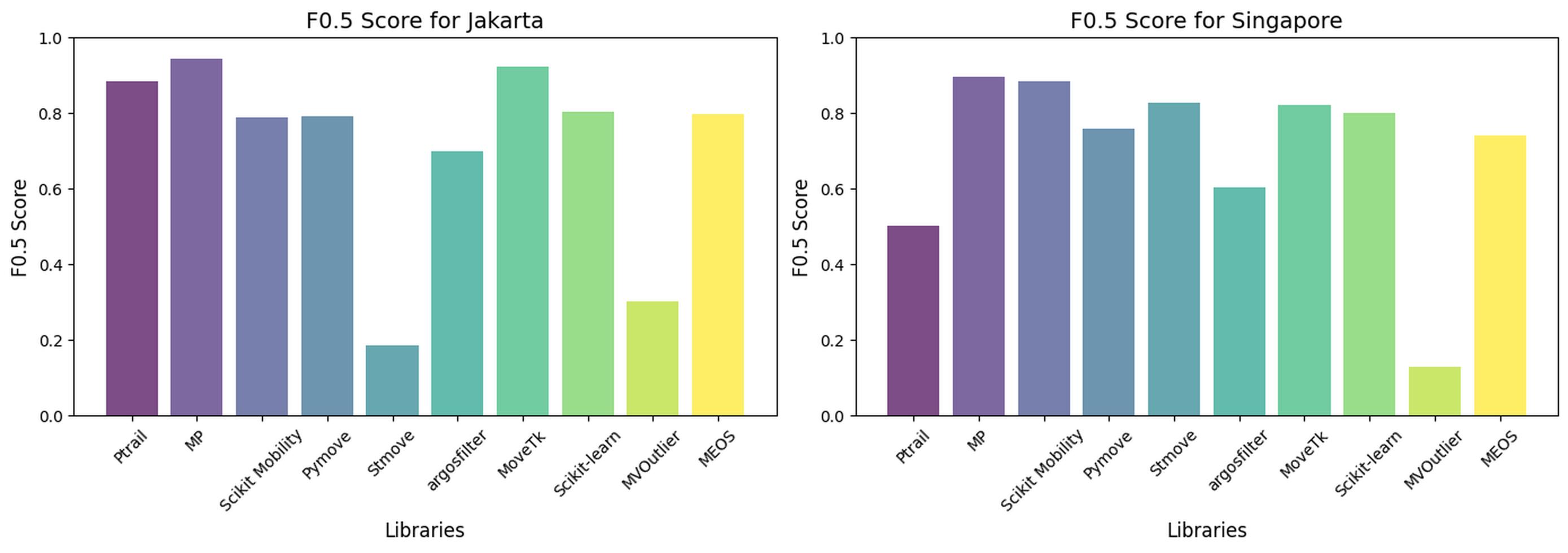}  
  \caption{F-0.5 score for libraries Ptrail, MovingPandas, Scikit-mobility, PyMove, Stmove, Argosfilter, MoveTK, Scikit-learn, MVOutlier and MEOS. The left graph corresponds to Jakarta datasets, the right graph represents the Singapore dataset.}
  \label{fig:LibrairesF0.5}
  \centering
   \includegraphics[width=0.85\linewidth]{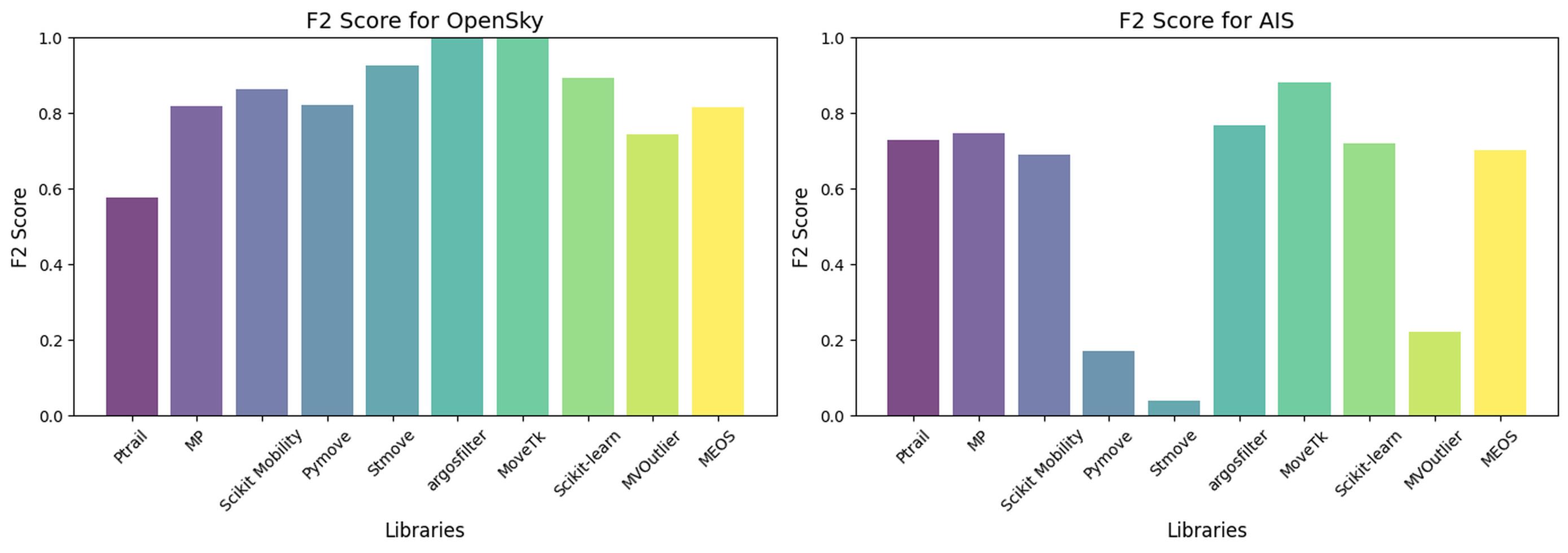}  
  \caption{F-2 score for libraries Ptrail, MovingPandas, Scikit-mobility, PyMove, Stmove, Argosfilter, MoveTK, Scikit-learn, MVOutlier and MEOS. The left graph represents the OpenSky dataset and the right is the AIS dataset}
  \label{fig:LibrairesF2}
\end{figure}

{\color{blue}
In order to analyze the correctness results, Figs.~\ref{fig:Precision},~\ref{fig:LibrairesRecall},~\ref{fig:LibrairesF1},~\ref{fig:LibrairesF2}, and~\ref{fig:LibrairesF0.5} show precision, recall, and F-1, F-2, and F-0.5 scores of each library, respectively. The scores are based on comparing the cross-check data and library output.

In Fig.~\ref{fig:Precision}, the precision is shown. The precision rate relates to the proportion of true positive predictions of outliers. This rate is homogeneous throughout libraries such as Ptrail, MovingPandas, and Sckit Mobility. Sckit Mobility outperforms all. Stmove presents the highest rate variation, especially for the AIS dataset.

In Fig.~\ref{fig:LibrairesRecall}, the metric of interest is the recall, which measures the ability of a library to identify true positives correctly, in this context, outliers within the dataset. A high recall score indicates that the library detects a large percentage of the true outliers, reducing the likelihood of missing significant outliers. Conversely, a low recall score suggests the library may be susceptible to false negatives, resulting in omitting non-outlier data points.
Of particular note is the performance of MoveTK, which displays commendable consistency across various datasets and emerges with the highest recall rate. Argosfilter and Scikit-learn closely follow this. On the other hand, MVOutlier and Ptrail exhibit significant variability in their recall rates, particularly underperforming in larger datasets. It suggests that both libraries offer less reliable outlier detection capabilities, a factor worth considering when selecting a library for specific data analysis requirements.
}

The F-1 score is a metric used for comparing the overall performance of the different libraries (Fig.~\ref{fig:LibrairesF1}). The MoveTk library has the biggest rate for the F1 score in all datasets. Followed by the Argos filter library. In the Jakarta dataset, Moving Pandas has the second-highest F-1 score. Followed by Argos filter and Pymove. In contrast, Stmove and Scikit Mobility have a low rate for the F-1 score. Moving Pandas has the third highest rate for the Singapore dataset, followed by StMove. Ptrail has the lowest rate in this dataset. All libraries have a similar performance in the OpenSky dataset, except for Ptrail, which has the lowest score in the first file. In the AIS dataset, Movetk, Ptrail, Moving Pandas, and Argos filter have the highest score. In contrast, Pymove and StMove have the lowest rate.

{\color{blue}To emphasize precision in our analysis, we have integrated the F0.5 score (Fig.~\ref{fig:LibrairesF0.5}), which holds greater weight in precision over recall. Conversely, the F2 score has been included to prioritize recall, shown in Fig.~\ref{fig:LibrairesF2}. The analysis of F0.5 and F2 scores across various libraries and datasets has yielded valuable insights into the performance of each library with a focus on precision and recall, respectively. When examining both F0.5 and F2 scores, we observe significant variability in library performance across different datasets, emphasizing the data-dependent nature of their effectiveness. Regarding F0.5 scores, Moving Pandas and Scikit Mobility are the top-performing libraries, excelling in precision, while MoveTk also demonstrated strong performance. However, when shifting the focus to F2 scores and prioritizing recall, Argos Filter and MoveTk are the top performers, indicating their effectiveness in capturing true outliers. These findings suggest that the library choice should be context-specific, considering the specific requirements of the task.

These results suggest that while Moving Pandas shows strong performance in precision, MoveTk excels in recall. Moving Pandas stands out with its high-precision
n (0.970) and strong performance in F-0.5 (0.920) and F-1 (0.842) scores, indicating its effectiveness in balancing recall and precision with a slight bias towards precision. On the other hand, MoveTk excels in recall (0.966) and F-1 score (0.923), suggesting its ability to identify true outliers. MEOS is a good choice, exhibiting balanced performance with F-0.5 (0.842), F-1 (0.752), F-2 (0.707), Precision (0.906), and Recall (0.673). Libraries like Scikit-learn and Argosfilter also demonstrate balanced capabilities, with Scikit-learn performing well in F-1 (0.851) and recall (0.862), and Argosfilter in F-2 (0.948) and recall (0.918). On the other hand, MVOutlier has lower scores in F-1 (0.179) and recall (0.301), implying limitations in its predictive accuracy. Despite their high precision, Ptrail and Pymove show moderate recall performance, indicating a potential trade-off between precision and recall in these libraries. }

\section{Conclusion}
\label{sec:conclusion}

In this paper, we evaluated the performance of ten libraries for outlier removal in trajectory data to guide data scientists and users on the existing libraries' offerings. The experimental study includes MovingPandas, Scikit-mobility, Scikit-learn, Ptrail, PyMove, movetk, MEOS, Argosfilter, Stmove and MOutlier. The datasets for this study consist of four massive datasets across multiple mobility domains, namely urban domain, air traffic, and marine. Our results suggest that the most reliable library is MoveTk. MoveTk presents a consistently high recall and F-1 score with the lowest run time. Scikit-learn follows closely, presenting strong performance metrics, thus making it another viable option for similar analytical tasks. Also, we have developed an approach for constructing ground-truth that involves cross-referencing the data from multiple sensors for speed and bearing. We applied this method to data from modern multi-sensor tracking technologies focusing on speed and bearing.

{\color{blue} It would be interesting to explore and experiment with algorithms implemented but not integrated within libraries for future work. Furthermore, apply methods from machine learning techniques to fine-tune the ground-truth method.}

\section*{Data and codes availability statement}
The data that support the findings of this study are available at GrabPosisi \cite{citiesData}, Denmark AIS data, \footnote{\url{http://web.ais.dk/aisdata/}} and OpenSky data.\footnote{\url{https://opensky-network.org/data/impala}} The codes are available in Github \footnote{\url{https://github.com/marianaGarcez/OutlierDetectionLibraries}}.

\bmhead{Acknowledgments}
This work was partially funded by the EU’s Horizon Europe research and innovation program under Grant No. 101070279 MobiSpaces.

\section*{Declarations}
The authors declare that they have no known competing financial interests or personal relationships that could have appeared to influence the work in this paper.

%\bibliography{sn-bibliography}

\begin{thebibliography}{44}
\providecommand{\natexlab}[1]{#1}
\providecommand{\url}[1]{{#1}}
\providecommand{\urlprefix}{URL }
\providecommand{\doi}[1]{\url{https://doi.org/#1}}
\providecommand{\eprint}[2][]{\url{#2}}
 \bibcommenthead

\bibitem[{Attia~Sakr and G{\"u}ting(2009)}]{attia2009spatiotemporal}
Attia~Sakr M, G{\"u}ting RH (2009) Spatiotemporal pattern queries in secondo.
  In: Advances in Spatial and Temporal Databases: 11th International Symposium,
  SSTD 2009 Aalborg, Denmark, July 8-10, 2009 Proceedings 11, Springer Berlin
  Heidelberg, pp 422--426

\bibitem[{Bakli et~al(2019)Bakli, Sakr, and Zimanyi}]{MobilityDB2}
Bakli M, Sakr M, Zimanyi E (2019) Distributed moving object data management in
  mobilitydb. In: Proceedings of the 8th ACM SIGSPATIAL International Workshop
  on Analytics for Big Geospatial Data, pp 1--10

\bibitem[{Breunig et~al(2000)Breunig, Kriegel, Ng, and Sander}]{LOF}
Breunig M, Kriegel HP, Ng R, et~al (2000) Lof: identifying density-based local
  outliers. In: Proceedings of the 2000 ACM SIGMOD international conference on
  Management of data, ACM, pp 93--104

\bibitem[{Brinkhoff(2002)}]{datasetwi}
Brinkhoff T (2002) A framework for generating network-based moving objects.
  GeoInformatica 6

\bibitem[{Cao et~al(2020)Cao, Liu, Meng, Liu, Miao, and Xu}]{window2}
Cao K, Liu Y, Meng G, et~al (2020) Trajectory outlier detection on trajectory
  data streams. IEEE Access PP:1--1

\bibitem[{Control(2022)}]{Europeanflights}
Control E (2022) The economics of aviation decarbonisation towards the 2030
  green deal milestone. Euro Control

\bibitem[{Custers et~al(2021)Custers, Kerkhof, Meulemans, Speckmann, and
  Staals}]{movetk}
Custers B, Kerkhof M, Meulemans W, et~al (2021) Maximum physically consistent
  trajectories. ACM Trans Spatial Algorithms Syst 7(4)

\bibitem[{Duarte and Sakr(2023)}]{DuarteS23}
Duarte M, Sakr M (2023) Outlier detection and cleaning in trajectories: {A}
  benchmark of existing tools. In: Proceedings of the Workshops of the
  {EDBT/ICDT} 2023 Joint Conference, Ioannina, Greece, March, 28, 2023, vol
  3379. CEUR-WS

\bibitem[{Eldawy and Mokhtar(2020)}]{Eldawy2020}
Eldawy E, Mokhtar H (2020) Clustering-based trajectory outlier detection.
  International Journal of Advanced Computer Science and Applications 11(5)

\bibitem[{Ester et~al(1996)Ester, Kriegel, Sander, and Xu}]{DBSCAN}
Ester M, Kriegel H, Sander J, et~al (1996) A density-based algorithm for
  discovering clusters in large spatial databases with noise. In: Proceedings
  of the Second International Conference on Knowledge Discovery and Data
  Mining. AAAI Press, KDD'96, p 226–231

\bibitem[{Filzmoser and Gschwandtner(2017)}]{mvoutlier}
Filzmoser P, Gschwandtner M (2017) {mvoutlier}: Multivariate outlier detection
  based on robust methods. R package

\bibitem[{Freitas et~al(2008)Freitas, Lydersen, Fedak, and
  Kovacs}]{argosfilter}
Freitas C, Lydersen C, Fedak MA, et~al (2008) A simple new algorithm to filter
  marine mammal argos locations. Marine Mammal Science

\bibitem[{Graser(2019)}]{movingPandas}
Graser A (2019) Movingpandas: Efficient structures for movement data in python.
  GI Forum Volume 7:54--68

\bibitem[{Haidri et~al(2021)Haidri, Haranwala, Bogorny, Renso, da~Fonseca, and
  Soares}]{haidri2021ptrail}
Haidri S, Haranwala YJ, Bogorny V, et~al (2021) Ptrail -- a python package for
  parallel trajectory data preprocessing. arXiv

\bibitem[{Huang et~al(2019)Huang, Yin, Lim, Wang, Hu, Varadarajan, Zheng,
  Bulusu, and Zimmermann}]{citiesData}
Huang X, Yin Y, Lim S, et~al (2019) Grab-posisi: An extensive real-life gps
  trajectory dataset in southeast asia. In: SIGSPATIAL, New York, NY, USA

\bibitem[{Jain(2010)}]{kmeans}
Jain A (2010) Data clustering: 50 years beyond k-means. Pattern Recognition
  Letters 31(8):651--666

\bibitem[{Knorr et~al(2000)Knorr, Ng, and Tucakov}]{Distance-BasedOutliers}
Knorr E, Ng R, Tucakov V (2000) Distance-based outliers: Algorithms and
  applications. The VLDB Journal 8:237--253

\bibitem[{Kotecha and Djuric(2003)}]{GparticleFilter}
Kotecha J, Djuric P (2003) Gaussian particle filtering. IEEE Transactions on
  Signal Processing 51(10):2592--2601

\bibitem[{Lee and West(2010)}]{particleFilter}
Lee SH, West M (2010) Performance comparison of the distributed extended kalman
  filter and markov chain distributed particle filter. IFAC Proceedings

\bibitem[{Magdy et~al(2017)Magdy, Sakr, and El-Bahnasy}]{MobilityDB4}
Magdy N, Sakr MA, El-Bahnasy K (2017) A generic trajectory similarity operator
  in moving object databases. Egyptian Informatics Journal 18(1):29--37

\bibitem[{{W}es {M}c{K}inney(2010)}]{Pandas}
{W}es {M}c{K}inney (2010) {D}ata {S}tructures for {S}tatistical {C}omputing in
  {P}ython. In: {S}t\'efan van~der {W}alt, {J}arrod {M}illman (eds)
  {P}roceedings of the 9th {P}ython in {S}cience {C}onference, pp 56 -- 61

\bibitem[{Moosavi et~al(2017)Moosavi, Omidvar-Tehrani, and Ramnath}]{anotation}
Moosavi S, Omidvar-Tehrani B, Ramnath R (2017) Trajectory annotation by
  discovering driving patterns. In: the 3rd ACM SIGSPATIAL Workshop

\bibitem[{Ng et~al(2002)Ng, Jordan, and Weiss}]{ng2002spectral}
Ng AY, Jordan MI, Weiss Y (2002) On spectral clustering: Analysis and an
  algorithm. In: Advances in neural information processing systems, pp 849--856

\bibitem[{Ng and Han(1994)}]{clarans}
Ng R, Han J (1994) Efficient and effective clustering methods for spatial data
  mining. In: Proceedings of the 20th International Conference on Very Large
  Data Bases. Morgan Kaufmann Publishers Inc., San Francisco, CA, USA, VLDB
  '94, p 144–155

\bibitem[{Oliveira(2019)}]{Pymove2}
Oliveira A (2019) Uma arquitetura e implementação do módulo de
  visualização para biblioteca pymove. Bachelor's thesis, UFC

\bibitem[{Pappalardo et~al(2019)Pappalardo, Simini, Barlacchi, and
  Pellungrini}]{scikit}
Pappalardo L, Simini F, Barlacchi G, et~al (2019) Scikit-mobility: a python
  library for the analysis, generation and risk assessment of mobility data

\bibitem[{Pearson et~al(2016)Pearson, Neuvo, Astola, and
  Gabbouj}]{hampelTheory}
Pearson R, Neuvo Y, Astola J, et~al (2016) Generalized hampel filters. EURASIP
  Journal on Advances in Signal Processing 2016

\bibitem[{Pedregosa et~al(2011)Pedregosa, Varoquaux, Gramfort, Michel, Thirion,
  Grisel, Blondel, Prettenhofer, Weiss, Dubourg, Vanderplas, Passos,
  Cournapeau, Brucher, Perrot, and Duchesnay}]{scikit-learn}
Pedregosa F, Varoquaux G, Gramfort A, et~al (2011) Scikit-learn: Machine
  learning in {P}ython. Journal of Machine Learning Research 12:2825--2830

\bibitem[{Sanches(2019)}]{Pymove1}
Sanches A (2019) Uma arquitetura e implementação do módulo de
  pré-processamento para biblioteca pymove. Bachelor's thesis, UFC

\bibitem[{Seidel et~al(2019)}]{stmove}
Seidel D, et~al (2019) Exploratory movement analysis and report building with r
  package stmove. bioRxiv

\bibitem[{Shi et~al(2021)Shi, Pan, Fang, and Chao}]{Shi2021RUTODRU}
Shi J, Pan Z, Fang J, et~al (2021) Rutod: real-time urban traffic outlier
  detection on streaming trajectory. Neural Computing and Applications
  35:3625--3637

\bibitem[{Thomas et~al(2017)Thomas, Barr, Balaji, and White}]{stoneSoup}
Thomas P, Barr J, Balaji B, et~al (2017) An open source framework for tracking
  and state estimation. In: Society of Photo-Optical Instrumentation Engineers
  (SPIE) Conference Series

\bibitem[{Trofficus(2021)}]{hampel}
Trofficus M (2021) Hampel filter in python

\bibitem[{Urrea and Agramonte(2021)}]{kf60}
Urrea C, Agramonte R (2021) Kalman filter: Historical overview and review of
  its use in robotics 60 years after its creation. Sensors

\bibitem[{Wang et~al(2019)Wang, Bah, and Hammad}]{newSurvey}
Wang H, Bah M, Hammad M (2019) Progress in outlier detection techniques: A
  survey. IEEE Access 7:107964--108000

\bibitem[{Wu et~al(2022)Wu, Zimanyi, Sakr, and Torp}]{song}
Wu S, Zimanyi E, Sakr M, et~al (2022) Semantic segmentation of ais trajectories
  for detecting complete fishing activities. In: 2022 23rd IEEE International
  Conference on Mobile Data Management (MDM). IEEE Computer Society

\bibitem[{Yang et~al(2022)Yang, Madsen, and Bednar}]{holoviz}
Yang S, Madsen M, Bednar J (2022) {HoloViz: Visualization and Interactive
  Dashboards in Python}. In: Proceedings of the 28th ACM SIGKDD Conference on
  Knowledge Discovery and Data Mining. SIGKDD

\bibitem[{Yang et~al(2018)Yang, Tang, and Li}]{datacleaning}
Yang X, Tang L, Li Q (2018) A data cleaning method for big trace data using
  movement consistency. In: Sensors

\bibitem[{Yu et~al(2017)Yu, Cao, Rundensteiner, and Wang}]{newslidingWindow}
Yu Y, Cao L, Rundensteiner E, et~al (2017) Outlier detection over massive-scale
  trajectory streams. ACM Trans Database Syst 42(2)

\bibitem[{Yuan et~al(2010)Yuan, Zheng, Zhang, Xie, Xie, Sun, and
  Huang}]{geopandas}
Yuan J, Zheng Y, Zhang C, et~al (2010) T-drive: Driving directions based on
  taxi trajectories. In: Proceedings of the 18th SIGSPATIAL International
  Conference on Advances in Geographic Information Systems. Association for
  Computing Machinery

\bibitem[{Zhang et~al(1996)Zhang, Ramakrishnan, and Livny}]{BIRCH}
Zhang T, Ramakrishnan R, Livny M (1996) Birch: An efficient data clustering
  method for very large databases. SIGMOD Rec 25(2):103–114

\bibitem[{Zheng et~al(2023)Zheng, Yu, Xie, and Wang}]{math11030620}
Zheng X, Yu D, Xie C, et~al (2023) Outlier detection of crowdsourcing
  trajectory data based on spatial and temporal characterization. Mathematics
  11(3)

\bibitem[{Zheng(2015)}]{SurveyTrajectoryDM}
Zheng Y (2015) Trajectory data mining: An overview. ACM Trans Intell Syst
  Technol 6(3)

\bibitem[{Zim\'{a}nyi et~al(2020)Zim\'{a}nyi, Sakr, and Lesuisse}]{MobilityDB}
Zim\'{a}nyi E, Sakr M, Lesuisse A (2020) Mobilitydb: A mobility database based
  on postgresql and postgis. In: ACM Trans. Database Syst., New York, NY, USA

\end{thebibliography}


\end{document}